\documentclass[twocolumn,english,pra,floatfix,showpacs,superscriptaddress,eqsecnum]{revtex4}
\usepackage[T1]{fontenc}
\usepackage[latin9]{inputenc}
\usepackage{babel}
\usepackage{amsmath}
\usepackage{amssymb}
\usepackage{graphicx}
\usepackage{esint}
\usepackage[unicode=true,pdfusetitle,
 bookmarks=true,bookmarksnumbered=false,bookmarksopen=false,
 breaklinks=false,pdfborder={0 0 1},backref=section,colorlinks=false]
 {hyperref}
\hypersetup{
 urlcolor=blue,hyperindex,colorlinks,linkcolor=black,citecolor=black}

\makeatletter

\DeclareRobustCommand{\greektext}{%
  \fontencoding{LGR}\selectfont\def\encodingdefault{LGR}}
\DeclareRobustCommand{\textgreek}[1]{\leavevmode{\greektext #1}}
\DeclareFontEncoding{LGR}{}{}
\DeclareTextSymbol{\~}{LGR}{126}
\newcommand{\lyxmathsym}[1]{\ifmmode\begingroup\def\b@ld{bold}
  \text{\ifx\math@version\b@ld\bfseries\fi#1}\endgroup\else#1\fi}

\@ifundefined{textcolor}{}
{%
 \definecolor{BLACK}{gray}{0}
 \definecolor{WHITE}{gray}{1}
 \definecolor{RED}{rgb}{1,0,0}
 \definecolor{GREEN}{rgb}{0,1,0}
 \definecolor{BLUE}{rgb}{0,0,1}
 \definecolor{CYAN}{cmyk}{1,0,0,0}
 \definecolor{MAGENTA}{cmyk}{0,1,0,0}
 \definecolor{YELLOW}{cmyk}{0,0,1,0}
 }

\usepackage{bbold}

\@ifundefined{definecolor}
 {\usepackage{color}}{}

\newcommand{\bra}[1]{\langle{#1}|}
\newcommand{\ket}[1]{|{#1}\rangle}

\newcommand{\be}{\begin{equation}}
\newcommand{\ee}{\end{equation}}
\newcommand{\bea}{\begin{eqnarray}}
\newcommand{\eea}{\end{eqnarray}}

\makeatother

\begin{document}

\title{Improving frequency selection of driven pulses using derivative-based
transition suppression}

\date{\today}

\author{F. Motzoi}

\affiliation{Institute for Quantum Computing and Department of Physics and Astronomy,
University of Waterloo, Waterloo, Ontario, Canada N2L 3G1}

\affiliation{Department of Chemistry, University of California, Berkeley, CA 94720
USA}

\author{F. K. Wilhelm}

\affiliation{Institute for Quantum Computing and Department of Physics and Astronomy,
University of Waterloo, Waterloo, Ontario, Canada N2L 3G1}

\affiliation{Theoretical Physics, Saarland University, 66123 Saarbrücken, Germany}
\begin{abstract}
Many techniques in quantum control rely on frequency separation as
a means for suppressing unwanted couplings. In its simplest form,
the mechanism relies on the low bandwidth of control pulses of long
duration. Here we perform a higher-order quantum-mechanical treatment
that allows for higher precision and shorter times. In particular,
we identify three kinds of off-resonant effects: i)~simultaneous
unwanted driven couplings (e.g. due to drive crosstalk), ii)~additional
(initially undriven) transitions such as those in an infinite ladder
system, and iii)~sideband frequencies of the driving waveform such
as we find in corrections to the rotating wave approximation. With
a framework that is applicable to all three cases, in addition to
the known\textbf{ }adiabatic error responsible for a shift of the
energy levels we typically see in the spectroscopy of such systems,
we derive error terms in a controlled expansion corresponding to higher
order adiabatic effects and diabatic excitations. We show, by also
expanding the driving waveform in a basis of different order derivatives
of a trial function (typically a Gaussian) these different error terms
can be corrected for in a systematic way hence strongly improving
quantum control of systems with dense spectra. 
\end{abstract}


\maketitle

\section{Introduction}

Spectroscopy is arguably the most commonly used experimental technique
in physics \cite{Slichter96,Demtroeder08,Levitt08}. It relies on
resonance - the object of study is exposed to monochromatic radiation
and responds if the radiation frequency matches a frequency of that
system. In quantum systems, that frequency is the difference of two
of the systems' energies. 

Complex systems usually contain a wealth of these frequencies. The
ability to selectively address these frequencies defines the spectral
resolution. The limitation of spectral resolution can be twofold:
on the one hand, the frequencies forming spectral lines are intrinsically
broadened by decoherence. On the other hand, an ideally monochromatic
external excitation is only a convenient fiction - in reality, the
bandwidth of that external signal is limited by a scale proportional
to $1/T$ where $T$ is the duration of the experimental pulse. In
magnetic resonance, e.g., certain spectral lines can only be reached
through complex pulse sequences that all need to be executed within
the relaxation time of the system. Consequently, a wealth of techniques
has been developed that reaches fine spectral selectivity with pulses
of limited duration, including 2D-spectroscopy\cite{Freeman98,Ernst90,Levitt08}
.

Quantum technologies such as quantum computing are often based on
spectroscopic ideas \cite{Chen06,Houches04,Nielsen00,DiV06}. In fact,
the already mentioned spin resonance is a primary candidate for the
implementation of quantum computing \cite{Gershenfeld97,qist,Baugh07}.
This means that the quantum mechanical transitions corresponding to
certain quantum logic operations are typically addressed through their
transition frequency. This can occur on the level of single qubits,
when the two states representing the qubit are taken out of a complex
spectrum with low anharmonicity such as it is the case in superconducting
qubits \cite{Devoret04,Insight,Makhlin01,Schoelkopf08,You05b}. It
can occur when multiple qubits are in close spatial proximity, much
closer than the spatial resolution of the external field, as it is
the case in spin resonance \cite{Freeman98,Levitt08,Nielsen00,Slichter96}.
It can also occur if single elements are multifunctional, e.g., when
a single qubit contains transitions pertinent to local rotations as
well as to coupling elements \cite{Blais07,Hofheinz09,Majer07,Plantenberg07,Sillanpaa07,Galiautdinov11}.
Examples of gate operations that contain transitions on single elements
are the NOT gate for single qubits and the controlled NOT and iSWAP
gates as multi-qubit operations \cite{Nielsen00}. Note that seriously
scalable quantum computing implementation candidates typically do
not rely on spectroscopic resolution alone and at least contain some
element of local addressability. Yet, clearly, a crowding of the frequency
spectrum will be detrimental to both spectroscopy and coherent quantum
control. 

In quantum information, it is a key requirement to perform a large
number of highly accurate operations well within the coherence time
of the system. Thus, the challenge of reaching good enough spectral
addressability in short times is of particular significance. Now a
key difference between spectroscopy and quantum control in the pursuit
of selective excitation is: Spectroscopy is an analytic technique
to find energy levels through transition frequencies, hence we want
to guarantee that beyond a narrow band around the desired transition,
excitation profiles are suppressed. In quantum control, the spectrum
is well characterized and the positions of undesired transitions are
known, hence, it is sufficient to suppress the excitation profile
at those frequencies. This paper aims primarily at the second approach.

Having a non-vanishing gap between energy levels is also the precondition
for applying the quantum adiabatic theorem. It turns out, as will
be made explicit later, that there is an equivalence between the spectral
excitation at an undesired transition and the inability to stay adiabatic
in the trajectory through parameter space taken by the controls. Several
studies have been undertaken to use the predictions of the adiabatic
theorem to avoid or to cancel the unwanted excitation \cite{Loy74,Unanyanc1997,Demirplak2003,Motzoi09,Berry09,Gambetta10,Guerin11,Torrontegui12,delCampo2012}.
In particular, Ref.~\cite{Demirplak2003} shows that including a
control operator to counter the diabatic error can emulate adiabatic
dynamics, and demonstrates how adding this (Lorentzian) control can
improve population transfer using chirped Adiabatic Passage techniques.
Furthermore, Ref.~\cite{Motzoi09} considers driving rotations on
a qubit whilst another transition nearby in frequency constitutes
leakage out of the qubit subspace. The result shows that one can simultaneously
rotate one transition while avoiding the other by using an off-phase
derivative of the driving waveform to cancel the diabatic error, allowing
for an adiabatic expansion of the joint dynamics, and was first verified
experimentally in Refs.~\cite{Chow10,Lucero10}. Ref.~\cite{delCampo2012}
considers removing the diabatic error when multiple homogenous transitions
are avoided for an Ising lattice. Ref.~\cite{Gambetta10} considers
the general case when multiple inhomogeneous transitions are present.
By using a Schrieffer-Wolff transformation the authors show how (in
principle) each order in the expansion can be optimized numerically
to minimize the aggregate error, in particular when using a constrained
set of controls.

In this paper, we further expand on these methods by constructing
analytical protocols for removing multiple unwanted transitions or
higher order errors. This is accomplished using a pseudo-adiabatic
expansion in a way that properly tracks the order of different types
of terms in the expansion. In effect, the technique generalizes the
Derivative Removal by Adiabatic Gate (DRAG) protocol of Refs.~\cite{Motzoi09,Gambetta10}
by including a set of higher-order derivatives. In the lowest order
of perturbation theory these constitute a basis with which a linear
set of equations approximating the differential equations giving the
effective spectrum of the waveform can be solved. Moreover, higher
order effects such as couplings to higher states in an anharmonic
ladder can be similarly removed using extra derivatives to satisfy
the additional constraints introduced by the higher-order effects. 

The paper is organized as follows: in Sec.~\ref{sec:selectivity},
we introduce the problem of selectivity and in Sec.~\ref{sec:fourieranalysis}
discuss it as an application of semi-classical sideband suppression,
deriving an asymptotic upper bound on off-resonant excitation related
to higher derivatives; in Sec.~\ref{sec:adiabatic}, we show how
the selectivity criteria can be derived for a quantum algebra and
define different ways to generalize it to multiple transitions; in
Sec.~\ref{sec:crosstalk} we apply the formalism to a set of frequency-separated
qubits and show how to use it to suppress crosstalk between them when
using a common drive. In Sec.~\ref{sec:Connected-transitions}, we
treat higher-orders in the problem of selective driving by considering
a ladder of connected transitions and show using higher derivatives
can prevent the (adiabatic) expansion from diverging. Sec.~\ref{sec:rabi}
discusses frequency selectivity and gives the example of very short
pulses where precise Rabi-like rotations can be maintained using the
same selectivity criteria.

\section{Quantum selectivity criteria \label{sec:selectivity}}

The controls that are used to manipulate quantum systems, typically
external AC fields, can often neither spatially nor by selection rule
distinguish between the quantum transition that is being controlled
and other quantum transitions. This can be mitigated if all these
transitions have distinct transition frequencies $ $$\omega_{j,k}=E_{k}-E_{j}$
where $E_{j}$ is the energy eigenvalue of state $j$ and here and
hereafter we use natural units with $\hbar=1$. If we now drive the
system control indexed by $l$ with a drive frequency $\omega_{l}^{d}$
that is much closer to a specific transition frequency labelled by
$j(l)$, $k(l)$ than to any other, and if this control has an appreciable
matrix element $\hat{\Gamma}_{jk}^{l}$ for this transition, only
it will be driven, and no other transition. We will quantify this
statement below and outline its limitations.

We start by assuming a Hamiltonian $\hat{H}_{\mathrm{sys}}=\hat{H}_{0}+\hat{H}_{{\rm control}}$
and work in the basis of eigenstates of $\hat{H}_{0}$. We can formalize
the statement about spectral selectivity by assuming that the drive
Hamiltonian has some appreciable matrix elements for multiple quantum
elements in the system, that is we have the control Hamiltonian 
\begin{equation}
\hat{H}_{{\rm control}}(t)=\sum_{l=0}^{p-1}\Omega_{l}(t)e^{-i\phi_{l}}\sum_{\{j,k\}}^{n}\hat{\Gamma}_{j,k}^{l}+\mathrm{h.c.,}\label{eq:basic_hamiltonian}
\end{equation}
where there are $n$ matrix elements (transitions) in the system and
$p$ drives to control them. 

As a typical example, this can arise if we consider $n$ qubits and
a collective drive composed of $p$ frequencies, each of which is
meant to address a particular qubit but has additional, unintended
crosstalk on the rest of the $n$ qubits, as it e.g. occurs in NMR.
Then the full Hamiltonian will read 

\begin{align}
\hat{H}_{\mathrm{control}}= & \sum_{l=1}^{p}2\Omega_{l}(t)e^{-i\phi_{l}}\cos\left(\int_{0}^{t}\omega_{l}^{d}(t^{\prime})dt^{\prime}\right)\sum_{m=1}^{n}\hat{\sigma}_{m}^{+}+\mathrm{h.c.}\label{eq:lineselec_lab-2}\\
\hat{H}_{0}= & \sum_{m=1}^{n}\frac{\omega_{0,1}^{m}}{2}(t)\hat{\sigma}_{m}^{z},\nonumber 
\end{align}
where we have left all terms time-dependent for generality. Other
examples, specifically where the transitions are not disjoint \textbf{(}i.e.,
they cannot be described using a tensor sum), will be discussed in
Secs. \ref{sec:Connected-transitions} and \ref{sec:rabi} and will
have similar forms. We can better appreciate the selectivity condition
by moving to the standard interaction picture, $\hat{H}_{I}$, and applying the rotating wave approximation, 
whereupon

\begin{equation}
\begin{split}
\hat{H}_{I} & = e^{-i\int H_0 dt}\hat{H}_{\mathrm{control}}e^{i\int H_0 dt}=\sum_l\sum_{j\neq k}\hat{\Gamma}_{j,k}^{l} \nonumber \\
\hat{\Gamma}_{j,k}^{l} & =  \lambda_{j,k}^{l}e^{-i\int_{0}^{t}\Delta_{jkl}(t')\mathrm{dt}'}|j\rangle\langle k|,\label{eq:drive_element}
\end{split}
\end{equation}
and the offsets $\Delta_{j,k,l}(t)=\omega_{l}^{d}(t)-\omega_{j,k}(t)$
define the distance from resonance of the transitions. The $\lambda_{j,k}^{l}$
weigh the relative strengths of the different transitions, letting
$\lambda_{j(l),k(l)}^{l}=1$. In the disjoint qubits example, we have
$|j\rangle=|0\rangle_{m}$, $|k\rangle=|1\rangle_{m}$. The evolution
of a system under the interaction Hamiltonian is then given by 

\begin{align}
U(0,T) & =\mathcal{T}\exp\left(\sum_{j,k,l}\int_{0}^{T}\Omega_{l}(t)e^{-i\phi_{l}}\hat{\Gamma}_{j,k}^{l}(t)\mathrm{dt}+\mathrm{h.c.,}\right),\label{eq:TimeEvol}
\end{align}
where $T$ is the evolution time and $\mathcal{T}$ enforces time-ordering.
Here, it is tacitly assumed that the envelope $\Omega_{l}={\rm Re}\Omega_{l}+i{\rm Im}\Omega_{l}$
is complex-valued. For the implementation of simple drive pulses,
the phase is typically assumed to be constant which with appropriate
choice of reference means ${\rm Im}\Omega=0$; however, later we will
explicitly use the ability to control both terms independently. 

Without loss of generality, we assume that to each driving field indexed
by $l$ we match a transition $j,k$ to which it is almost resonant,
identified as $j(l)$ and $k(l)$.\emph{ }In this interaction frame
representation, we can formulate the sufficient conditions for selectivity:
in order for the drive element Eq.~\ref{eq:drive_element} to be
effective, it must oscillate more frequently than the time scale of
the transition $\left|\lambda_{j,k}^{l}\Omega\right|{}^{-1}$ but
less than $\left|\lambda_{j(l),k(l)}^{l}\Omega\right|{}^{-1}$, specifically
\begin{align}
\Delta_{j,k,l} & \gg\left|\lambda_{jk}^{l}\Omega_{l}\right|\quad\forall j,k\not=j(l),k(l)\label{eq:selectivity_inequality}\\
\Delta_{j(l),k(l),l} & \ll\left|\lambda_{j(l),k(l)}^{l}\Omega_{l}\right|\label{eq:resonance_condition}
\end{align}
 As $\omega_{j,k}$ are given by the quantum system under consideration,
the choice of driving frequencies $\omega_{l}$ can only maximize
the left hand side of Eq. \ref{eq:selectivity_inequality} to a certain
limit set by the need to obey Eq. \ref{eq:resonance_condition}. Thus,
obeying these conditions requires to keep the control amplitudes $\Omega_{l}$
low enough, which increases the duration of the control pulse, but
makes the transition vulnerable to decoherence and relaxation. Thus,
we practically demand that $ $$\lambda_{j(l),k(l)}^{l}\Omega_{l}\gg\gamma$
where $\gamma$ represents typical incoherent rates of the system.
This constraint on addressability is a result of spectral crowding
and the loss of fidelity due to the need for long pulses degrades
spectroscopic techniques as well as the implementation of coherent
gates in a quantum computer. Thus, the spectrum sets a speed and fidelity
limit on quantum control. We will derive these conditions and bound
them using the Fourier transform in the next section.

\section{Asymptotic limit \label{sec:fourieranalysis}}

\subsection{Fourier transform}

It is well established\cite{Freeman98,Hoult79,Vold68,Warren84} that
for a system of qubits or spins 1/2 driven by a weak external field,
i.e., for small $\Omega/\Delta$, an accurate measure of off-resonant
excitation is the Fourier transform 

\begin{equation}
\begin{split}S(\Omega,\Delta) & =\int_{0}^{T}\Omega(t)e^{-i\Delta t}dt\end{split}
\label{-3}
\end{equation}
That is, at long times and large frequency separation $\Delta$, the
time-ordering terms in Eq. \ref{eq:TimeEvol} will commute and the
time-ordering operator can be dropped \cite{Warren84}. Note that
this is a limited-interval Fourier transform that can be consolidated
with the regular, infinite-time Fourier transform by assuming the
pulse envelope $\Omega(t)$ vanishes outside the integration interval.
The conditions for selectivity (Eqs. \ref{eq:selectivity_inequality}
and \ref{eq:resonance_condition}) then imply that, for large times,

\begin{equation}
\begin{split}S(\Omega,\Delta_{j,k,l}) & =\theta\delta_{j,j(l)}\delta_{k,k(l)}\end{split}
\label{eq:FTselcond}
\end{equation}
where $\delta_{a,b}$ is the Kronecker delta. When the time-ordering
can be dropped \cite{Haeberlen68}, Eq. \ref{eq:TimeEvol} then gives
back trivially a $\theta$ rotation on (only) the desired transition
$j-k$. 

A caveat to this approximation is that off-resonant levels will induce
additional phase errors (coming from, e.g., AC Stark shifts) for shorter
times or multiple drives, coming from enforcing time-ordering. In
practice, these can be corrected by some combination of adjusting
resonance conditions, applying compensatory gates to undo the accumulated
phase at the end of the operation, or by inserting frame transformations
between operations (see Appendix A). The derivation of these phase
terms will be discussed in detail in later sections when we consider
the full dynamics of concrete examples (see also \cite{Gambetta10}).

One well-established way to compensate spectral weight off-resonance
while still maintaining a pulse of limited length is to use pulse
shaping \cite{Freeman98}. For this purpose, it is customary \cite{Bauer1984,Steffen2003}
to use Gaussian profiles, which are well-confined Gaussians both in
the frequency and time domains. In this case, the Gauss function describing
$\Omega(t)$ must be suitably chosen to start and end at zero amplitude
and takes the form

\begin{equation}
\Omega_{G}(t)=A\left(\exp\left[-\frac{(t-T/2)^{2}}{2\sigma^{2}}\right]-\exp\left[-\frac{(T/2)^{2}}{2\sigma^{2}}\right]\right)^{m}.\label{eq:gaussian}
\end{equation}
Here, $\sigma$ is the standard deviation, $m$ is chosen such that
$m-1$ derivatives of the function start and end at $0$, and $A$
is chosen such that the correct amount of rotation is implemented
(e.g. $A=\pi/\sqrt{2\pi\sigma^{2}}\mathrm{erf}[T/\sqrt{8}\sigma]$
for an area $\pi$ pulse). We will follow a different strategy that,
rather than \emph{reducing }off-resonant excitations for a full \emph{band}
of energies, \emph{eliminates }excitation for one or more $discrete$
frequencies.

\subsection{Order counting\label{sub:ordercounting}}

For this purpose, we wish to be able to quantify and remove the effect
of unwanted off-resonant error. To be able to remove multiple such
errors, we will want to find equivalent formulations of the error,
which we will see below will be found using different orders of differentiation
of the driving waveform. Finally, we will want to see the effect polynomial
functions of these derivatives as these will be needed if we want
to Taylor expand the dynamics (Sec. \ref{sec:adiabatic}) in terms
of these different orders. Having established the role of the Fourier
transform, we can now adapt an idea from classical calculus and signal
processing. We start from the excitation profile for detuning $\Delta$,
then integrate by parts (IBP), assume that the envelope and its lowest
$n$ derivatives vanish in both the beginning and the end of the pulse,
and find

\begin{align}
S(\Omega,\Delta) & =\int_{0}^{T}\Omega(t)e^{-i\Delta t}\mathrm{dt}\nonumber \\
 & =-i\int_{0}^{T}\frac{\frac{d}{dt}\Omega(t)}{\Delta}e^{-i\Delta t}\mathrm{dt}\label{eq:IBP}\\
 & =(-i)^{n}\int_{0}^{T}\frac{\frac{d^{n}}{dt^{n}}\Omega(t)}{\Delta^{n}}e^{-i\Delta t}\mathrm{dt}\nonumber 
\end{align}
This result tells us that the spectral weight of the $n$-th derivative
of the control signal will be amplified by a factor $\Delta^{n}$
relative to the original waveform, or asymptotically will be $\Theta\left(S(\Delta^{n}\Omega(t),\Delta)\right)$.
Moreover, this equivalence will hold over infinitesimal intervals
$[t,t+dt]$ of the full evolution as well (neglecting the boundary
terms, which will cancel between intervals). Likewise it is easy to
see that derivatives of polynomial functions of the waveform will
obey the same formula $S\left(\frac{d^{m}}{dt^{m}}\left(\prod_{k}(\frac{d^{k}\Omega}{dt^{k}})^{n_{k}}\right),\Delta\right)=\Theta(S(\Delta^{m}\prod_{k}(\frac{d^{k}\Omega}{dt^{k}})^{n_{k}},\Delta))$.
More generally, it can also be verified numerically that, asymptotically
in $\Delta$, 

\begin{equation}
S\left(\left(\Omega(t)\right)^{\sum n_{k}},\Delta\right)=\mathcal{O}\left(S\left(\prod_{k}\left(\frac{d^{k}\Omega(t)}{dt^{k}}\right){}^{n_{k}},\Delta\right)\right)\label{eq:asymptbehaviour}
\end{equation}
for given $n_{k}$.

In the quantum limit, when in the adiabatic regime ($\Omega<\Delta$),
we will see in the next sections that adiabatic expansions around
a small parameter $\epsilon=\frac{\lyxmathsym{\textgreek{W}}}{\text{\ensuremath{\Delta}}}$
will obey the same infinitesimal-time asymptotics, as commutators
of terms in $S(\epsilon,\Delta)$ will be of the next or higher order
in $\epsilon$ and hence not contribute. In addition, in the extreme
limit ($\Omega\ll\Delta$), the full integral over $T$ will commute
with other small terms and thereby accurately predict off-resonant
excitation (since the Fourier transform is a good measure). We will
see in Sec. \ref{sec:adiabatic} how these terms can be accounted
for to precisely calculate gate errors for shorter times as well.

\subsection{Engineering of the instantaneous spectrum\label{sub:semiclassicaldrag}}

\begin{figure}
\centering  \includegraphics[width=0.25\textwidth]{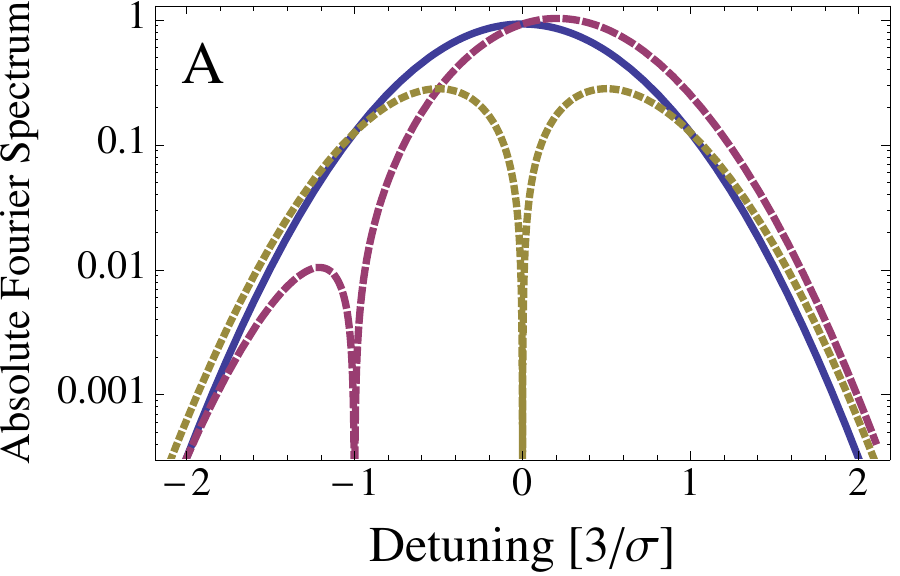}\includegraphics[bb=0bp 0bp 235bp 147bp,width=0.21\textwidth]{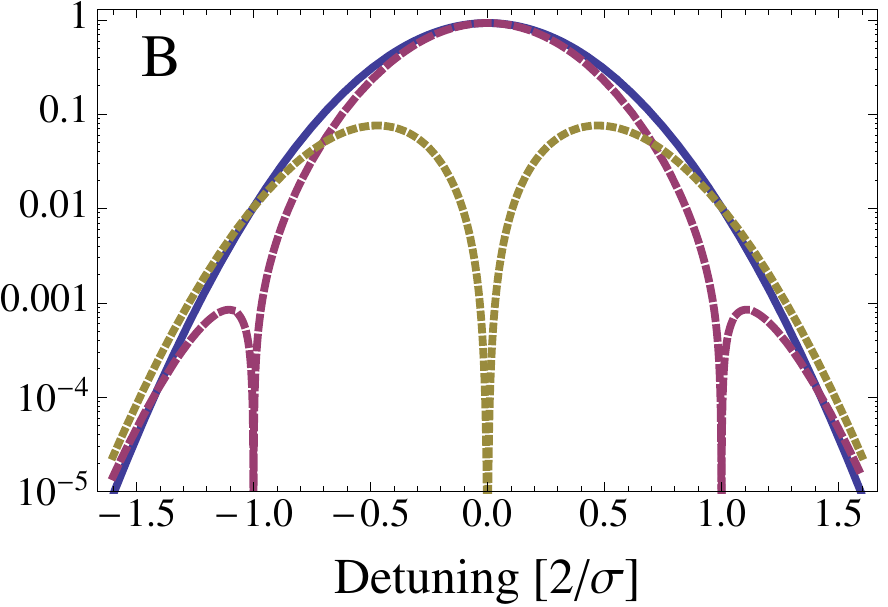}

\caption[Absolute Fourier transform of a Gaussian pulse (solid blue), its derivative
(dotted yellow), and their sum (dashed red). (A) is the first derivative;
(B) is the second derivative.\emph{ }]{Absolute Fourier transform of a Gaussian pulse, its derivative, and
their sum. (A) is the first derivative (with $\sigma=3$); (B) is
the second derivative (with $\sigma=2$). }
\label{fig:dragspect}
\end{figure}

For now, we describe a semi-classical strategy for utilizing the correspondence
between the frequency spectra of the waveform and its derivatives
to suppress off-resonant excitation. The strategy is to satisfy Eqs.
\ref{eq:FTselcond} by supplementing the waveform with some small
auxiliary controls proportional to the derivatives,

\begin{align}
\Omega(t) & =\mathrm{Re}\Omega(t)+i\mathrm{Im}\Omega(t)\label{eq:generalDragFirstOrder}\\
 & =\Omega_{0}(t)+\sum_{r=1}^{n/2}a_{2r}\frac{d^{2r}}{dt^{2r}}\Omega_{0}(t)+i\sum_{r=1}^{n/2}b_{2r-1}\frac{d^{2r-1}}{dt^{2r-1}}\Omega_{0}(t)\nonumber 
\end{align}
where $a_{i}$ and $b_{i}$ will be chosen to satisfy the selectivity
constraints. For example, using a Gaussian as our base waveform $\Omega_{0}(t)=\Omega_{G}(t)$,
we can engineer a hole in the spectrum at a frequency offset $\Delta$
from the driven transition. The simplest way to do this is by choosing
\begin{equation}
{\rm Im}\Omega(t)=-\frac{{\rm Re}\dot{\Omega}(t)}{\Delta},\label{eq:drag}
\end{equation}
i.e., $b_{1}=-\frac{1}{\Delta}$. The spectrum of this control shape
is illustrated in Fig. \eqref{fig:dragspect}A. The zero of the frequency
axis is set to the wanted transition, the undesired transition is
placed at $\Delta=-3/\sigma$ ($\sigma$ being the standard deviation
of the Gaussian). The Gaussian definitely has appreciable spectral
weight at the unwanted transition. The derivative also has spectral
weight there, so the difference with the appropriate weight (-$\small\frac{i}{\Delta}$)
will be zero. By construction, the derivative of the Gaussian has
no spectral weight at the working transition (it is anti-symmetric)
hence does not alter the spectral profile of the working transition.
Note that the perturbation caused by the auxiliary control is small
both in the time and frequency domains as it suppressed by a factor
$\Delta^{-1}$. Thus, we see the derivatives of the function have
two effects: on the one hand they result in a disproportionately large
error off-resonance (Eq. \ref{eq:IBP}); but on the other, with the
introduction of a small perturbation we are able to completely cancel
out the undesired excitation from the principle waveform (Eq. \ref{eq:generalDragFirstOrder}).

We can apply the same technique for higher-order derivatives. For
example, the second-derivative solution $a_{2}=\frac{1}{\Delta^{2}}$,

\begin{align*}
\Omega(t)= & \Omega_{0}(t)+\frac{1}{\Delta^{2}}\ddot{\Omega}_{0}(t)
\end{align*}
 will satisfy Eqs. \ref{eq:FTselcond} provided ${\rm \int_{0}^{T}}\Omega(t)\mathrm{dt}=\theta$.
The effect is demonstrated in Fig.~\ref{fig:dragspect}B. Off resonance,
unwanted transitions are cancelled at the chosen $\Delta=\pm2/\sigma$
while maintaing full rotation on resonance. We choose $\Omega_{0}(t)=\Omega_{G}(t)$
with $m=2$ to ensure the IBP formula is valid twice over in Eq. \ref{eq:IBP}.
The strategy may be preferable to the first derivative solution in
certain cases. Since the first derivative is anti-symmetric it increases
excitation at $\Delta=+1$, thus the second derivative may be more
useful when transitions are not wanted on both sides of resonance,
as would happen for a spectrum with a Liouvillian degeneracy, i.e.,
with distinct transitions having equal frequencies. Moreover, we can
see the overall bandwidth (above a given signal-to-noise threshold--
here 0.001) is decreased compared to a traditional Gaussian by about
25\%, where some of the energy has been moved from the selective region
to the tails where it instead falls below threshold. This could be
useful when a continuum of excitations needs to be avoided, as in
resonance spectroscopy. Lastly, using only controls in phase with
each other implies they commute (they obey an area theorem, integrating
to $\theta$) avoiding higher order effects such as phase shifts and
rotation errors on the working transition. 

Finally, let us notice again that since Eq. \ref{eq:IBP} holds for
infinitesimal times as well, the spectrum engineering is far more
effective than simply obeying Eq. \ref{eq:FTselcond}. For example,
the pulse shape $\Omega(t)=\Omega_{0}(t)+\frac{1}{\Delta}e^{i\Delta(t-T)}\dot{\Omega}_{0}(t)$$ $
will also have the same average spectrum at the critical frequencies
and benefit from being only a small perturbation; however, this solution
is not valid over intervals smaller than $T$ ($S(\Omega(t),\Delta)$
does not vanish for small time intervals), and hence the time-ordering
operator in Eq. \ref{eq:TimeEvol} cannot be easily accounted for.
Thus, in what follows, we will only consider instantaneous-time solutions
such as we have found above, which will allow for an instantaneous-time
expansion of the dynamics in Sec. \ref{sec:adiabatic}.

\subsection{Multiplet engineering\label{sub:Semiclassical-multiplet-suppress}}

We can generalize the semi-classical solution to suppressing multiple
unwanted excitations. Specifically, for crowded spectra and high precision
requirements, it may not be sufficient to only put one or two holes
in the spectrum and rely on bandwidth constraints for the rest. Instead,
we must now solve Eq. \ref{eq:FTselcond} for multiple offsets, $\{\Delta_{j}\}$.
Plugging Eq. \ref{eq:generalDragFirstOrder} into Eq. \ref{eq:FTselcond}
and applying the IBP formulae Eq \ref{eq:IBP}, we null the integrand
to obtain

\begin{align}
1+\sum_{r=1}^{n/2}(-1)^{r}(\Delta_{j})^{2r}a_{2r}-\sum_{r=1}^{n/2}(-1)^{r}(\Delta_{j})^{2r-1}b_{2r-1} & =0,\label{eq:multidrag}
\end{align}
where $n$ is the number of undesired transitions. Such a system of
linear equations can easily be solved. For example, for $n=2$, the
structure of the solution is

\[
\Omega(t)=\Omega_{0}(t)-i\left(\frac{1}{\Delta_{1}}+\frac{1}{\Delta_{2}}\right)\dot{\Omega}_0(t)+\frac{\ddot{\Omega}_{0}(t)}{\Delta_{1}\Delta_{2}}
\]
or alternatively

\[
\Omega(t)=\Omega_{0}(t)+\frac{\Delta_{1}^{3}+\Delta_{2}^{3}}{\Delta_{1}^{3}\Delta_{2}-\Delta_{1}\Delta_{2}^{3}}i\dot{\Omega}_0(t)+\frac{i\dddot{\Omega}_{0}(t)}{\Delta_{1}^{2}\Delta_{2}-\Delta_{1}\Delta_{2}^{2}}.
\]
For $n=3$ we can use

\begin{align*}
\Omega(t)= & \Omega_{0}(t)-i\left(\frac{1}{\Delta_{1}}+\frac{1}{\Delta_{2}}+\frac{1}{\Delta_{3}}\right)\dot{\Omega}_0(t)+\\
 & \frac{\Delta_{1}+\Delta_{2}+\Delta_{3}}{\Delta_{1}\Delta_{2}\Delta_{3}}\ddot{\Omega}_{0}(t)+\frac{1}{\Delta_{1}\Delta_{2}\Delta_{3}}i\dddot{\Omega}_{0}(t).
\end{align*}

While these semi-classical solutions will be valid in the limit of
$\Omega\ll\Delta$, they will become less accurate as higher-order
derivatives are used, as is typical for an asymptotic expansion. It
will be crucial to supplement them with corrections to rotation angle
and resonance/phase errors and other higher-order quantum effects.

\section{Frequency-selective adiabatic expansion\label{sec:adiabatic}}

In the limit of $\Omega\ll\Delta$, the primary effect of the driving
field is to time-dependently change the energies in the system. This
can be understood as an application of the adiabatic theorem, or to
higher order, the super-adiabatic expansion \cite{Lim1991}. On the
other hand, we have seen with Eq. \ref{eq:IBP} that in exactly the
same regime, the excitation of unwanted transitions will occur proportionally
to derivatives of the waveform, and we will refer to this error as
diabatic error. Our goal will be to remove both the diabatic and the
adiabatic errors, which inhibit perfect rotation of the working qubit,
in particular when multiple unwanted transitions exist in the system.
In this section, we will show how the dynamics can be expanded in
terms of a small parameter to compute and suppress these errors, in
particular in extension to previous works by considering the expansion
in terms of the derivatives (which we will see naturally arise in
the adiabatic expansion). We will be able to characterize and suppress
the order-of-magnitude diabatic effect of terms involving derivatives
relative to the small parameter by using the asymptotic scaling found
in Sec. \ref{sub:ordercounting}. 

We choose to work in a (computational) frame where the time-independent
part of the Hamiltonian has been diagonalized (into, in general, dressed
eigenstates) so that energy transitions are clearly defined by the
difference in diagonal entries in the Hamiltonian matrix. We then
perform a sequence\textbf{ }of time-dependent transformations that
allow us to obtain instantaneous-time control operators for which
gate synthesis is trivial.

\subsection{Rotating frame\label{sub:rotatingframe}}

We start by moving to a frame where all transitions between adjacent
energy levels are rotating at the frequency of the drive. For clarity,
we choose indices so that adjacent levels have minimal energy difference
between them. Our goal, once the frame transformation has been performed,
is that the matrix elements corresponding to the drive for adjacent
levels will contain a term that does not oscillate (and whose only
time-dependence comes from that of the drive waveform envelope). These
elements will constitute the primary error with respect to selectively
driving a particular transition. Other less significant terms, including
counter-rotating terms, matrix elements between non-adjacent levels,
and extra drive terms used to simultaneously drive other transitions,
will oscillate at their sideband frequency relative to the rotating
frame. Sec. \ref{sec:rabi} gives an example of how these errors can
be suppressed using the same formalism. 

The rotating frame transformation is defined by

\[
R=\exp\left(-i\sum_{k=1}^{N}\int_{0}^{t}\Delta_{k-1,k,l}(t)\mathrm{dt}\ket{k}\bra{k}\right)
\]
and the transformed frame with respect to the interaction picture
given by 

\begin{align}
\hat{H}_{R}= & R\hat{H}_{I}R^{\dagger}+i\dot{R}R^{\dagger}\nonumber \\
= & \sum_{l=0}^{L-1}\Omega_{l}(t)e^{-i\phi_{l}}\sum_{\{j,k\}}^{n+1}\lambda_{j,k}^{l}e^{-i\int_{0}^{t}\omega_{j,k-1}(t)\mathrm{dt}}\ket{j}\bra{k}+\mathrm{h.c.}\label{eq:RotatingFrame}\\
 & +\sum_{k=1}^{N}\Delta_{k-1,k,l}(t)\ket{k}\bra{k}\nonumber 
\end{align}
where $\Delta_{j,k,l}=\omega_{l}^{d}-\omega_{j,k}$ is defined as
before. For the simple case outlined above then we have in particular
$\omega_{j,k-1}=0$ and the only time-dependence comes from $\Omega_{l}(t)$.
As a final preparation step, we separate the rotating Hamiltonian
into wanted, unwanted, and irrelevant (e.g. diagonal) terms. Thus,
we can write equivalently

\begin{align*}
\hat{H}_{R}= & \hat{H}_{w}^{\mathrm{}}+\hat{H}_{uw}+\hat{H}_{\mathrm{diag}}
\end{align*}
where

\begin{align*}
\hat{H}_{w} & =\Omega_{l}(t)e^{-i\phi_{l}}\ket{j(l)}\bra{k(l)}\\
\hat{H}_{uw} & \approx\Omega_{l}(t)e^{-i\phi_{l}}\sum_{j,k}^{n}\lambda_{j,k}^{l}\ket{j}\bra{k}+\mathrm{h.c.\quad\forall j,k\not=j(l),k(l)}\\
\hat{H}_{\mathrm{diag}} & =\sum_{k=1}^{N}\Delta_{k-1,k,l}(t)\ket{k}\bra{k}.
\end{align*}
Note that we have neglected oscillating terms for clarity in this
final form.

\subsection{Block diagonal frame\label{sec:singlet}}

From Eq. \ref{eq:RotatingFrame}, we now want to move to another frame
where the remaining time-dependence can be trivially calculated. To
do this we will find a diagonal representation for the unwanted transitions
so that only an (easily correctable) phase shift will result on these
levels. The problem lies with being able to find such a diagonalizing
transformation. A general transformation of this kind will be of the form

\[
\hat{H}_{D}=D\hat{H}_{R}D^{\dagger}+i\dot{D}D^{\dagger}.
\]
In general, for fast pulses, the second term on the right hand side
can have a larger contribution than the terms that were being diagonalized,
as we have seen in Sec. \ref{sub:ordercounting}. Instead, we will
use intuition from Sec. \ref{sub:semiclassicaldrag} to define an
interaction frame with respect to an auxiliary (off-phase, derivative)
control operator, within which the diagonalization will be well-defined.
That is

\begin{align}
\hat{H}_{\mathrm{tot}} & =(\hat{H}_{w}+\hat{H}_{uw}+\hat{H}_{\mathrm{diag}})+\hat{H}_{\mathrm{aux}}\label{eq:dragexact}\\
D & =\exp\left(i\int_{0}^{t}\hat{H}_{\mathrm{aux}}(t)\mathrm{dt}\right)\nonumber \\
\widetilde{H} & =D(\hat{H}_{w}+\hat{H}_{uw}+\hat{H}_{\mathrm{diag}})D^{\dagger}\nonumber \\
 & =\widetilde{H}_{w}\oplus\widetilde{H}_{\mathrm{diag}}.\nonumber 
\end{align}
Thus, $\hat{H}_{\mathrm{aux}}$ is chosen such that $D$ diagonalizes
$\hat{H}_{uw}$ provided $\hat{H}_{\mathrm{aux}}$ is also chosen
such that it averages to zero over the time $T$ and that it commutes
with itself at different times. The final form can be written more
explicitly as 

\begin{equation}
\begin{split}\widetilde{H}_{w} & =\tilde{\Omega^{{\rm }}}(t)e^{-i\phi}\ket{j(l)}\bra{k(l)}+\mathrm{h.c.}\\
\widetilde{H}_{uw} & =0\\
\int_{0}^{T}\tilde{\Omega^{{\rm }}}(t)dt & =\theta,
\end{split}
\label{eq:quantselcond}
\end{equation}
which is the quantum mechanical equivalent of Eq. \ref{eq:FTselcond}.
This technique can be used to exactly solve certain transition selection/avoidance
problems, such as the two-qubit crosstalk problem in Sec. \ref{sec:doublet}.

In general, it is not possible to analytically diagonalize $\hat{H}_{R}$
exactly and instead one will have to use perturbation theory with
respect to a small parameter, here chosen as 
\begin{equation}
\epsilon=\max_{jk}\left(\frac{|\lambda_{jkl}|\Omega(t)}{\Delta_{jkl}}\right).\label{eq:smallparam}
\end{equation}
 In effect, if more than one unwanted transition elements exist in
the system then one will need more than the single auxiliary control
waveform to cancel out the undesired dynamics. That is, one will need
to solve a system of differential equations relating to the diagonalization
transformation and time derivative of the transformation for each
of the unwanted off-diagonal elements (Eq. \ref{eq:quantselcond}).
Once again, we can use the intuition from Sec. \ref{sub:Semiclassical-multiplet-suppress}
to solve the system using a basis of higher order derivatives as an
ansatz. Let the controls be defined analogously to Eq. \ref{eq:generalDragFirstOrder}
by

\begin{align}
\Omega(t) & =\mathrm{Re}\Omega+i\mathrm{Im}\Omega=\Omega_{0}+\sum_{r=1}^{n}\Omega_{r}^{Q}+i\sum_{r=1}^{n}\Omega_{r}^{I}\nonumber \\
\hat{H}_{\mathrm{aux}} & =\sum_{r}\hat{H}_{r}^{Q}+\sum_{r}\hat{H}_{r}^{I}\nonumber \\
\hat{H}_{r}^{Q} & =\Omega_{r}^{Q}(t)e^{-i\phi+i\pi}\sum_{j,k}\lambda_{j,k}^{l}\ket{j}\bra{k}+\mathrm{h.c.}\label{eq:controldecomp}\\
\hat{H}_{r}^{I} & =\Omega_{r}^{I}(t)e^{-i\phi}\sum_{j,k}\lambda_{j,k}^{l}\ket{j}\bra{k}+\mathrm{h.c.}\nonumber 
\end{align}
Each of the $n$ unwanted transitions will be approximately diagonalized
by a combination of a real and imaginary operator, thus there are
$2n$ operators that define the (self-commuting for different times)\textbf{
}transformations

\begin{align}
D(t) & =\exp\left(-i\hat{Y}\right)=\exp\left(i\sum_{r}\hat{Y}_{r}\right)\label{eq:transfwithderivatives}\\
E(t) & =\exp(-i\hat{X})=\exp\left(i\sum_{r}\hat{X}_{r}\right)\nonumber 
\end{align}
where the real and imaginary operators have been applied in separate
transformations and contain higher-derivative contributions. Applying
the transformations in sequence gives the effective Hamiltonian

\begin{align}
\widetilde{H}= & D\left(E\hat{H}_{\mathrm{tot}}E^{\dagger}\right)D^{\dagger}+iD\dot{E}E^{\dagger}D^{\dagger}+i\dot{D}D^{\dagger}\label{eq:multipletransitionsdrag}\\
= & D\left(E(\hat{H}_{\mathrm{diag}}+\hat{H}_{\mathrm{aux}})E^{\dagger}+\hat{H}_{w}+\hat{H}_{\mathrm{uw}}\right)D^{\dagger}\nonumber \\
 & +D\dot{\hat{X}}D^{\dagger}+\dot{\hat{Y}}\nonumber \\
= & \widetilde{H}_{w}\oplus\bigoplus_{r}^{n-1}\widetilde{H}_{(r)}^{\mathrm{diag}}\nonumber 
\end{align}
The last line will hold only if the transformations combine to time-independently
diagonalize the unwanted transitions ($j-k$) of the Hamiltonian, that is
\begin{align}
\tilde{H}_{jk} & =0,\label{eq:QuantEqSys}
\end{align}
\textbf{$ $}for each $j,k\neq j(l),k(l).$ The simultaneous diagonalization of
these unwanted transitions sets up a system of equations. Remarkably, the linearization of this system (Eqs.~\ref{eq:transfwithderivatives} - \ref{eq:multipletransitionsdrag}) in terms of $\hat{X}_{r}$ and $\hat{Y}_{r}$ is identical to the semiclassical system, Eq.~\ref{eq:multidrag}.
More concretely, we have the linear approximation

\begin{align}
\left(\hat{H}_{\mathrm{uw}}+\sum_{r}^{n}([i\hat{Y}_{r},\hat{H}_{\mathrm{diag}}]+[i\hat{X}_{r},\hat{H}_{\mathrm{diag}}]\label{eq:QuantDragLinearization}\right.\\
\left. +\hat{H}_{r}^{Q}+\hat{H}_{r}^{I}+\dot{\hat{X}}_{r}+\dot{\hat{Y}}_{r})\right)_{jk} & \approx0,\nonumber 
\end{align}
which must hold for each pair \{$j,k$\} not amongst the desired transition(s).
Thus, starting with the unwanted transition element $\hat{(H}_{\mathrm{uw}})_{jk}$,
each successive order of transformation diagonalizes the unwanted
off-diagonal transition element (first line), and leaves in its place a derivative
that is only \emph{partially} canceled by the auxiliary derivative
control $\hat{H}_{r}^{Q(I)}$ of that order (second line). This process is iterated
with each order leaving a higher-order, off-diagonal derivative with
a smaller prefactor. By the $n$-th order, the off-diagonal contribution
has been fully removed in aggregate by all the auxiliary controls,
and ensuring this happens for all unwanted pairs \{$j,k$\} solves
the system of equations. 

Moreover, Eq. \ref{eq:multipletransitionsdrag} allows $\tilde{H}_{w}$
and $\tilde{H}_{\mathrm{diag}}$ to, be computed, unlike the semiclassical
approach where these are assumed to stay at their bare values. Variations
in $\tilde{H}_{w}$ from the ideal Eq. \ref{eq:quantselcond} will
result in resonance and rotation errors on the working transition.
Variations in $\tilde{H}_{\mathrm{diag}}$ will be a little more subtle
and will primarily result in time-dependently changing values of \{$\Delta_{j}$\}
in Eq. \ref{eq:multidrag}. Last but not least, higher order in $\epsilon$
will result not only in corrections to Eq. \ref{eq:QuantDragLinearization}
but also in new (unwanted) transitions $j-k$ not amongst the $n$
transitions initially driven in the bare frame, Eq. \ref{eq:RotatingFrame}. 

Note that the system of linear differential equations, Eq. \ref{eq:multipletransitionsdrag},
has been replaced by a system of (linear) algebraic equations, Eq.
\ref{eq:QuantDragLinearization}, where only the multiplicative factor
in front of the derivatives need be computed. That is, the derivatives
form a basis for the evolution of the populations, and thus the prefactors
in front of them that solve the algebraic equations also solve the
differential equation at all time. Thus, the result is an instantaneous
solution which solves the (unwanted excitation) problem exactly at
all times in the evolution. This is similar to the conclusion at the
end of Sec. \ref{sub:Semiclassical-multiplet-suppress}, but now it
holds to higher orders, including the fact that resonance and rotation
errors that are only uncovered in the quantum mechanical treatment
and do not occur in the semiclassical order can also be computed and
suppressed. An example to suppressing multiple undesired transitions
is given in Sec. \ref{sec:crosstalk}, which demonstrates solutions
to the multi-qubit crosstalk problem.

When we further include the non-linear (higher-order) effects in the
small parameter $\epsilon$ (defined in Eq. \ref{eq:smallparam})
of the evolution, we will no longer be able to completely cancel terms
containing derivatives, as cross-terms will arise that contain more
than a single order of differentiation of the trial function (e.g.
$\dot{\Omega}^{2}\ddot{\Omega}$). Nonetheless, going to higher orders
may be necessary for high-precision control or to apply further error-correcting
protocols. Qualifying the effect of these terms will be important
as the number of error terms will grow exponentially with each order
in the expansion (though in practice very few orders will be needed
to suppress the error). Moreover, expansions that calculate the effect
of these terms such as adiabatic or super-adiabatic expansions will
typically diverge \cite{Lim1991,Deschamps2008} and so knowing which
terms cause divergence will be crucial to controlling the expansion
and avoiding divergences. These systematic effect of these terms can
be gauged (and reconciled with the previous paragraph) by considering
that the $k$-th order in $\epsilon$ will contain at most $\frac{n!}{k!}$
different cross-terms, and these will form a basis for the evolution
to that order. To do this we expand around the semi-classical, large
$T$ limit (that is, the adiabatic limit) given in Sec.\ref{sub:ordercounting}
so that we collect terms up to each order $\mathcal{O}(\epsilon^{k}$)
in the expansion. Instead of Eq. \ref{eq:multipletransitionsdrag},
the transformation is identified recursively up to each order by

\begin{align}
\hat{H}(t) & =\tilde{H}^{(0)}(t)=\hat{H}_{R}+\sum_{g}\hat{H}_{\mathrm{g,0}}^{({\rm aux})}\nonumber \\
\widetilde{H}^{(h)}(t) & =\hat{D}_{h}(t)\tilde{H}^{(h-1)}(t)\hat{D}_{h}^{\dagger}(t)+i\dot{\hat{D}}_{h}(t)\hat{D}_{h}^{\dagger}(t)\nonumber \\
 & =\widetilde{H}_{w}^{(h)}\oplus\widetilde{H}_{\mathrm{diag}}^{(h)}+\mathcal{O}(\epsilon^{k})\nonumber \\
\tilde{H}_{\mathrm{g},h}^{(\text{{\rm aux}})}(t) & =\hat{D}_{h}(t)\tilde{H}_{\mathrm{g},h-1}^{({\rm aux})}\hat{D}_{h}^{\dagger}(t)\label{eq:dragformalism}
\end{align}
where $h$ indexes the order of the frame transformation, $g$ gives
the (derivative) order of the auxiliary controls, and $k$ indexes
the order of the error for the given frame. Since certain transitions
will not correspond to controlled transitions (amongst the original
$n$), not all transformations $D_{h}$ will correspond to an interaction
frame and hence $h\geq k$. For the ones that do, we once again have 

\[
D_{h}(t)=\exp\left(i\int_{0}^{t}\tilde{H}_{\mathrm{k},h}^{(\mathrm{aux})}(t)\mathrm{dt}\right)
\]
In other cases, one simply has the diagonalization

\[
D_{h}(t)=\exp\left(i\hat{S}_{h}(t)\right).
\]
Then, plugging these into Eq.~\ref{eq:dragformalism} and taking the k-th order expansion of the Baker-Campbell-Hausdorff lemma, the effective $h-$th order
Hamiltonians are defined as

\begin{align}
\widetilde{H}^{(h)}= & H^{(h-1)}+i[\hat{H}_{k,h}^{T},\tilde{H}^{(h-1)}]\label{eq:iterativeInteraction}\\
 & +\frac{1}{2}[\hat{H}_{k,h}^{T},[\hat{H}_{k,h}^{T},\tilde{H}^{(h-1)}]]+\mathcal{O}(\epsilon^{k}),\nonumber 
\end{align}
where $\hat{H}_{k,h}^{T}=\int_{0}^{t}\tilde{H}_{k,h}^{(\mathrm{aux})}(t')\mathrm{dt'}$,
and

\begin{align}
\widetilde{H}^{(h)}= & H^{(h-1)}+i[\hat{S}_{h},\tilde{H}^{(h-1)}],\label{eq:IterativeAdiabatic}\\
 & +\frac{1}{2}[\hat{S}_{h},[\hat{S}_{h},\tilde{H}^{(h-1)}]]+\dot{\hat{S}}_{h}+\mathcal{O}(\epsilon^{k}),\nonumber 
\end{align}
respectively. The goal is to pick auxiliary controls $\hat{H}_{\mathrm{g,0}}^{(aux)}$
such that in the higher-order transformed frame they cancel out with
unwanted excitation error (note the lack of derivative in Eq. \ref{eq:iterativeInteraction}, $i\dot{\hat{D}}_{h}(t)\hat{D}_{h}^{\dagger}(t)=-\tilde{H}_{\mathrm{k},h}^{(\mathrm{aux})}$).
Similarly, $\hat{S}_{h}$ can be calculated as is typically done with
the Schrieffer-Wolff transformation \cite{Schrieffer1966}, essentially
the diagonalizing operator to the next order (but introducing a higher derivative of the same order of error). As in the other cases,
the transformation corresponds to an time-dependent, adiabatic one
(now an adiabatic expansion) within the interaction frame given by
the auxiliary controls. Note that outside the interaction frame, the
evolution is not strictly adiabatic as the states followed by the
system are not eigenstates of the complete Hamiltonian as is typically
true of the adiabatic theorem. In fact, we may not even be in the
adiabatic regime at all as the derivatives may be large. In the following
three sections we go through examples for suppressing multiple unwanted
transitions, higher order errors, and sideband frequency errors.

\section{Disjoint transitions\label{sec:crosstalk}}

We demonstrate the formalism for the multiple qubit problem, with
the simplification that couplings between the qubits are neglected
for our purposes. In this situation, the primary error when driving
single qubit rotations will be unwanted crosstalk to other qubits
from the external driving field. The lab frame Hamiltonian is given
by Eq. \ref{eq:lineselec_lab-2}, which we transform to the rotating
frame as shown in Sec. \ref{sub:rotatingframe} to give

\begin{align}
\hat{H}_{R}= & \Omega(t)e^{-i\phi_{l}}\sum_{m=1}^{n}\lambda_{m}\hat{\sigma}_{m}^{+}+\mathrm{h.c.}\label{eq:disjointqubits}\\
 & +\sum_{m=1}^{n}\Delta_{m}(t)\ket{1}\bra{1}_{m}\nonumber 
\end{align}
with $\Delta_{j}=\omega_{0}^{d}(t)-\omega_{0,1}^{(j)}$.

\subsection{Exact doublet solution\label{sec:doublet}}

The simplest scenario is for two qubits for which we can use Eq. \ref{eq:dragexact}
to remove crosstalk exactly. We put the drive on resonance with the
first qubit (detuning away from $\Delta_{1}(t)=0$ will be used later
to cancel phase error) and let $\Delta_{2}(t)=\Delta_{1}(t)+\Delta(t)$.
Using the transformation

\begin{align}
D= & \Biggl(\cos\Bigl(\int_{0}^{t}e^{-i\phi}{\mathrm{Im}}\Omega(t)\lambda_{1}dt\Bigr)\hat{{\mathbb{1}}}\nonumber \\
 & \mathrm{+\sin\Bigl(\int_{0}^{t}e^{-i\phi}{\mathrm{Im}}\Omega(t)\lambda_{1}dt\Bigr)\hat{\sigma}_{1}^{+}}+\mathrm{h.c.}\Biggr)\\
 & \otimes\Biggl(\cos\Bigl(\int_{0}^{t}e^{-i\phi}{\mathrm{Im}}\Omega(t)\lambda_{2}dt\Bigr)\hat{{\mathbb{1}}}\label{eq:quanttransf-1}\\
 & \mathrm{+\sin\Bigl(\int_{0}^{t}e^{-i\phi}{\mathrm{Im}}\Omega(t)\lambda_{2}dt\Bigr)\hat{\sigma}_{2}^{+}}+\mathrm{h.c.}\Biggr),
\end{align}
 and solving Eq. \ref{eq:quantselcond} for qubit 2 subspace then
gives the solution

\begin{equation}
e^{-i\phi}\lambda_{2}\int_{0}^{t}{\mathrm{Im}}\Omega(t)dt=\frac{1}{2}\tan^{-1}\left(\frac{2e^{-i\phi}\lambda_{2}\mathrm{{Re}\Omega(t)}}{\Delta}\right)\label{eq:dragquant}
\end{equation}
to avoid crosstalk, or equivalently

\begin{equation}
{\mathrm{Im}}\Omega(t)=\frac{\Delta\mathrm{{Re}\dot{\Omega}(t)}}{\Delta^{2}+(2\lambda_{2}\mathrm{{Re}\Omega(t))}^{2}}.\label{eq:dragquant-1}
\end{equation}
This is the quantum mechanical version of the IBP formula, Eq. \ref{eq:drag}.
In particular its first order Taylor expansion is the same, and can
often be easier to work with. This solution also bears close resemblance
to the diabatic error term in the adiabatic theorem, which would be
\[
H_{\mathrm{diab}}=\frac{i\lambda_{2}e^{-i\phi}\dot{\Omega}(t)}{\sqrt{\Delta^{2}+(2\lambda_{2}\mathrm{{Re}\Omega(t))}^{2}}}\sigma_{2}^{+}+\mathrm{h.c.}
\]
in the absence of a perturbation (the denominator is the instantaneous-time
energy). The discrepancy occurs on account of the perturbation also
introducing diabatic error, and hence needing itself to be corrected. 

\begin{figure}
\centering \includegraphics[width=0.8\columnwidth]{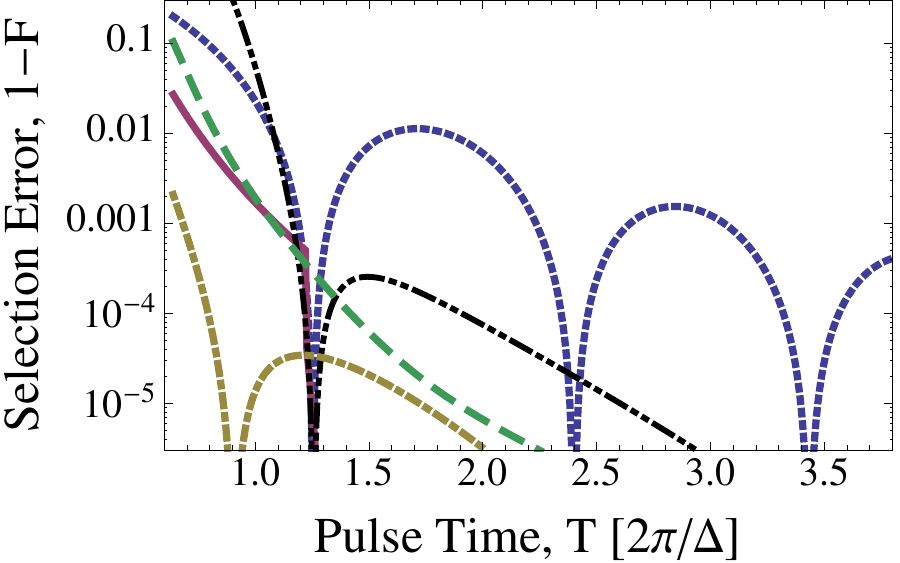} \caption{Population inversion error (Eq. \ref{eq:selectionerror}) for two
uncoupled qubits of energy difference $\Delta$. The dotted blue line
shows the error using standard Gaussian shaping, Eq. \ref{eq:gaussian},
while the solid red line uses the pulse shape given by DRAG Eqs. \ref{eq:dragquant}-\ref{eq:dragdetun},
which is an exact (infinite-order) solution for $T>2.5\pi/\Delta$. The dot-dashed orange
line is the error when using the first-order shape given by the second-derivative
solution, Eq. \eqref{eq:2ndDerive_2qubits}; the dashed green line
corresponds to the first-order shape using the third derivative (Eq.
\eqref{eq:3rdDerive_2qubits}); and the dot-dot-dashed black line
is for the first-order, fourth-derivative solution, Eq. \eqref{eq:4thDerive_2qubits}.
\label{fig:Population-selection-(inversion)}$ $}
\end{figure}
Introducing an imaginary part of the control will also affect the
dynamics of the first qubit. Solving again Eq. \ref{eq:quantselcond}
we get

\begin{eqnarray}
 &  & \Delta_{1}(t)=-2\mathrm{Re}\Omega(t)\tan\left(2\int_{0}^{t}\mathrm{Im}\Omega(t')\mathrm{dt'}\right)\label{eq:dragdetun-1}\\
 &  & \int_{0}^{T}\mathrm{Re}\Omega(t)\sec\left(2\int_{0}^{t}\mathrm{Im}\Omega(t')\mathrm{dt}'\right)dt=\theta\nonumber 
\end{eqnarray}
For Eq. \ref{eq:dragquant-1} with $\lambda_{2}=1$, these take on
the simple form

\begin{eqnarray}
 &  & \Delta_{1}(t)=\frac{1}{2}\left(-\text{\ensuremath{\Delta}}+\sqrt{\text{\ensuremath{\Delta}}^{2}-(4{\mathrm{Re}}\Omega(t))^{2}}\right)\label{eq:dragdetun}\\
 &  & \int_{0}^{T}\frac{\Omega(t)|\Delta|}{\sqrt{\Delta_{2}(t)^{2}+4({\mathrm{Re}}\Omega(t))^{2}}}=\theta\nonumber 
\end{eqnarray}

To quantify the selection error, we use the phase-insensitive quantum
fidelity for a unitary map, given by 

\begin{equation}
F=\frac{1}{2^{n}}\mathrm{Tr}|U^{\dagger}V|,\label{eq:selectionerror}
\end{equation}
where $U$ is the evolution given by the chosen set of controls, and
$V$ is the desired evolution, here $\hat{\sigma}_{x}\otimes^{n-1}\hat{\mathbb{\mathbf{\mathbb{1}}}}$
(with $n=2)$. The selection error ($1-F$) is plotted as a function
of gate time in Fig.~\ref{fig:Population-selection-(inversion)}.
The error for a simple Gaussian of correct area is plotted in dotted
blue, while the DRAG solution is in solid red. The Gaussian does suppress
the selection error at certain times, where Eqs. \ref{eq:quantselcond}
are all satisfied (or approximately Eqs. \ref{eq:FTselcond}, see
Ref. \cite{Warren84}), though clearly the pulses are susceptible
to pulse-time miscalibrations. On the other hand, the Gaussian pulse
with DRAG correction is an exact solution at all times beyond about
$2.5\pi/\Delta$, that is, when $4\Omega(t)<\Delta_{2}$ is met. At
shorter times the detuning (Eq. \ref{eq:dragdetun}) has no real solution,
so the first order solution is plotted instead, still outperforming
the Gaussian.

\subsection{General class of solutions\label{sec:quantderivs}}

For more qubits, Eqs. \ref{eq:multipletransitionsdrag} must be solved.
For two-level systems, the transformation Eq. \ref{eq:transfwithderivatives}
can be parametrized with $\hat{X}=\sum_{j}X_{j}\sigma_{j}^{+}$, $\hat{Y}=\sum_{j}Y_{j}\sigma_{j}^{+}$such
that

\begin{align}
D(t) & =\exp\left(i\sum_{j=1}^{M}Y_{j}\hat{\sigma}_{j}^{+}-h.c\right)=\exp\left(i\sum_{j,r}Y_{j,r}\hat{\sigma}_{j}^{+}-h.c\right)\label{eq:transfwithderivatives-1}\\
E(t) & =\exp\left(i\sum_{j=1}^{M}X_{j}\hat{\sigma}_{j}^{+}-h.c\right)=\exp\left(i\sum_{j,r}X_{j,r}\hat{\sigma}_{j}^{+}-h.c\right),\nonumber 
\end{align}
where the index $r$ refers to orders of derivatives of the trial
function and $j$ indexes the transition. These simplify the constraints
to solve for (Eq. \ref{eq:QuantEqSys}), giving a system of $2n$
equations 

\begin{eqnarray*}
(\mathrm{Re}\Omega-\dot{X_{j}})\cos(2Y_{j})-\frac{1}{2}\Delta_{j}\text{\ensuremath{\cos}}(2X_{j})\sin(2Y_{j})\\
+\mathrm{Im}\Omega\sin(2X_{j})\sin(2Y_{j}) & = & 0\\
-i\mathrm{Im}\Omega\cos(2X_{j})+i\frac{1}{2}\Delta_{j}\sin(2X_{j})+i\dot{Y_{j}} & = & 0
\end{eqnarray*}
which in turn gives rise to the self-consistency equation 

\begin{eqnarray}
\sin2X_{j} & = & \frac{1}{\Delta_{j}}\frac{\frac{d}{dt}\left(\frac{\mathrm{Re}\Omega-\dot{X_{j}}}{\Delta_{j}\cos(2X_{j})+\mathrm{Im}\Omega\sin(2X_{j})}\right)}{\frac{1}{2}+\left(\frac{\mathrm{Re}\Omega-\dot{X_{j}}}{\Delta_{j}\cos(2X_{j})+\mathrm{Im}\Omega\sin(2X_{j})}\right)^{2}}\label{eq:selfconsistency}\\
 &  & -2\mathrm{Im}\Omega\cos(2X)\nonumber 
\end{eqnarray}
$ $ We decompose the control waveform $\Omega(t)$ into a series
of derivatives as per Eq. \ref{eq:controldecomp}. The self-consistency
equation can be solved exactly using iterative solutions. In this
section, rather, it will suffice to linearize the equations in terms
of the derivatives,

\begin{equation}
X_{j}=\frac{\lambda_{j}\sum_{r}\mathrm{\dot{\Omega}_{r}^{Q}(t)-\sum_{r}\ddot{X}_{j,r}}}{\Delta_{j}^{2}+(2\lambda_{j}\mathrm{Re}\Omega(t))^{2}}-\frac{\mathrm{\lambda_{j}\sum_{r}}\Omega_{r}^{I}}{\Delta_{j}}\label{eq:crosstalkequations}
\end{equation}
which must be solved for all $\{\Delta_{j}\}$. To obtain a simple
formula for crosstalk removal, in analogy to the simple equations,
Eq. \ref{eq:multidrag}, we can approximate the prefactors of each
derivative control (Eq. \ref{eq:generalDragFirstOrder}) as approximately
time independent and thus obtain

\begin{align}
1+\sum_{r=1}^{N/2}(-1)^{r}(\Delta_{j}^{2}+(2\lambda_{j}\Omega_{0})^{2})^{r}a_{2r}\label{eq:multidrag-1-1}\\
-\sum_{r=1}^{N/2}\frac{(-1)^{r}(\Delta_{j}^{2}+(2\lambda_{j}\Omega_{0})^{2})^{r}}{\Delta_{j}}b_{2r-1} & =0,\nonumber 
\end{align}
 In addition, we must worry about resonance and rotation errors on
the working qubit, which will be significant. These can be exactly
unwound again using \ref{eq:dragdetun-1}.

We have set up with Eqs. \ref{eq:multidrag-1-1} and \ref{eq:dragdetun-1}
a framework for independent control of two-level systems via a common
drive. The main difference compared to the IBP formula Eq. \ref{eq:generalDragFirstOrder}
comes from consolidating the frequency offset with the time-dependent
energies of the unperturbed qubits, $\sqrt{\Delta_{j}^{2}+(2\lambda_{j}\Omega_{0}(t))^{2}}$.
Eq. \ref{eq:dragquant-1} gives the exact relation. In the subsections
that follow, we solve for explicit forms of the solutions to the systems
for 2, 3, and 4 qubits. The result is straightforwardly generalized to
larger systems, but the results are cumbersome to display.

\subsubsection{Doublet solutions }

To demonstrate this class of solutions with the general form Eq. \ref{eq:crosstalkequations},
consider again the two-qubit system (Eq. \ref{eq:disjointqubits}
with $n=2)$. Since the system contains only one transition we want
to cancel, we can solve for it directly. In addition to the first
derivative solution given by Eq. \ref{eq:dragquant-1}, a different
solution exists for each higher derivative. For instance, the second
derivative solution, for which by construction $\mathrm{Im}\Omega=0$,
can be found by setting $\Omega_{2}^{Q}=\dot{X}$ in Eq. \ref{eq:crosstalkequations},
from which

\begin{align}
\Omega(t) & =\Omega_{0}+\frac{d}{dt}\frac{\dot{\Omega}_{0}}{\Delta^{2}+(2\lambda_{2}\Omega_{0})^{2}}.\label{eq:2ndDerive_2qubits}
\end{align}
The selection error (Eq. \ref{eq:selectionerror}) for this pulse
sequence is plotted in dot-dashed orange in Fig.~\ref{fig:Population-selection-(inversion)}
, clearly out-performing the Gaussian result as well as the exact,
first-derivative solution in the very short time regime where the
adiabaticity of the eigenstates breaks down, $\dot{\Omega}>\Delta(t)$.
To general order, choosing the recurrence relation $X_{i}=\frac{-1}{\Delta^{2}+4(\lambda_{2}\mathrm{Re}\Omega(t))^{2}}\ddot{X}_{i-1}$,
gives the (approximate) general real and imaginary solutions

\begin{eqnarray*}
\Omega_{r}^{R} & = & \frac{d}{dt}\prod_{q=1}^{r/2}\left(\frac{1}{\Delta^{2}+(2\lambda_{2}\Omega_{0}(t))^{2}}\frac{d^{2}}{dt^{2}}\right)\frac{\mathrm{\dot{\Omega_{0}}(t)}}{\Delta^{2}+(2\lambda_{2}\mathrm{\Omega_{0}(t))}^{2}}\\
\end{eqnarray*}

\begin{eqnarray*}
\Omega_{r}^{I} & = & \prod_{q=1}^{\frac{r-1}{2}}\left(\frac{1}{\Delta^{2}+(2\lambda_{2}\Omega_{0}(t))^{2}}\frac{d^{2}}{dt^{2}}\right)\frac{\Delta\mathrm{\dot{\Omega_{0}}(t)}}{\Delta^{2}+(2\lambda_{2}\mathrm{\Omega_{0}(t))}^{2}}
\end{eqnarray*}
The fourth derivative real solution is

\begin{align}
\Omega(t) & =\Omega_{0}(t)+\Omega_{4}^{Q}(t)\label{eq:3rdDerive_2qubits}\\
 & =\Omega_{0}(t)+\frac{d}{dt}\frac{1}{\Delta^{2}+(2\lambda_{2}\Omega_{0}(t))^{2}}\frac{d^{2}}{dt^{2}}\frac{\mathrm{\dot{\Omega_{0}}(t)}}{\Delta^{2}+(2\lambda_{2}\mathrm{\Omega_{0}(t))}^{2}}.\nonumber 
\end{align}
The third derivative complex solution is given by

\begin{equation}
\Omega_{3}^{I}(t)=\frac{\Delta}{\Delta^{2}+(2\lambda_{j}\Omega_{0}(t))^{2}}\frac{d^{2}}{dt^{2}}\frac{\mathrm{\dot{\Omega_{0}}(t)}}{\Delta^{2}+(2\lambda_{2}\mathrm{{Re}\Omega(t))}^{2}}\label{eq:4thDerive_2qubits}
\end{equation}
The controls being complex, the working transition will also be affected
but we can unwind the detuning and rotation error exactly using Eq.
\ref{eq:dragdetun}. The selection errors for the two pulse sequences
are plotted in Fig.\ref{fig:Population-selection-(inversion)} as
the dashed green line for the third derivative and the dot-dot-dashed
black line for the fourth derivative, still outperforming the Gaussian
result in the long-time (adiabatic) limit.

The same methodology can be followed to obtain solutions involving
even higher order derivatives. It is important to emphasize that in
order for the effective frame to be equivalent to the original bare
frame we must choose a pulse shape $\Omega_{0}(t)$ whose derivatives
are zero at the endpoints of the pulse (e.g. Eq. \ref{eq:gaussian}).

\subsubsection{Triplet solutions}

\begin{figure}
\centering \includegraphics[width=0.8\columnwidth]{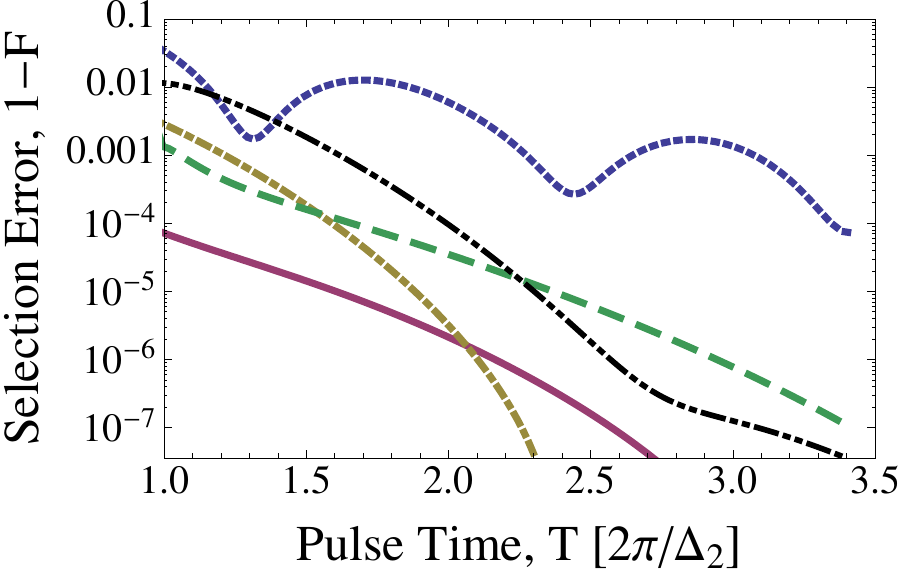} \caption[Selection error for three uncoupled qubits for use of different derivatives]{Selection error (Eq. \ref{eq:selectionerror}) for three uncoupled
qubits for energy differences from the first qubit of $\Delta_{2}$
and $\Delta_{3}=1.7\Delta_{2}$. The dotted blue line shows the combined
inversion error using standard Gaussian shaping, Eq. \ref{eq:gaussian}.
The solid red line gives the error for the solution using the first
and second derivatives, Eq. \eqref{eq:1stand2nd_3qubits}; the dot-dashed
orange is for the first and third derivatives, Eq. \eqref{eq:1stand3rd_3qubits};
the dashed green line is for the second and third derivatives, Eq.
\eqref{eq:2ndand3rd_3qubits}; and the dot-dot-dashed black line is
for the second and fourth derivatives, Eq. \eqref{eq:2ndand4th_3qubits}.
\label{fig:Selection-error-for}}
\end{figure}

If more than 2 qubits are in the system, with only a single amplitude
control it becomes increasingly difficult (at the cost of larger $T$)
to find a gate where crosstalk is avoided on all other qubits using
a Gaussian pulse. The recurrences of low errors in Fig. \ref{fig:Population-selection-(inversion)}
roughly every $2$ units of time become suppressed or disappear when
additional crosstalk qubits are included because the underlying Bohr
frequencies become incommensurate. For example, adding a qubit at
$\Delta_{3}=1.7\Delta_{2}$, we see in Fig.~\ref{fig:Selection-error-for}
that the Gaussian pulse (in dotted blue) no longer performs nearly
as well as it did for two qubits (Fig. \ref{fig:Population-selection-(inversion)}).

Thus, it is all the more useful to apply the higher-derivative formalism
to suppress crosstalk for more qubits. The various solutions are plotted
in Fig.~\ref{fig:Selection-error-for} for the 3 qubit problem. Eqs.
\ref{eq:multidrag-1-1} can be straightforwardly solved by hand for
small $n$, while the task is aided by a computer algebra system as
the formulae will become cumbersome. For the first and second derivative
solution, we get 
\begin{eqnarray}
\Omega(t) & = & \Omega_{0}(t)+ib_{1}(t)\dot{\Omega}_{0}(t)+\frac{d}{dt}a_{2}(t)\dot{\Omega}_{0}(t),\label{eq:1stand2nd_3qubits}\\
 & = & \Omega_{0}-i\frac{\Delta_{2}\Delta_{3}\left(E_{2}^{2}-E_{3}^{2}\right)}{E_{2}^{2}E_{3}^{2}\left(\Delta_{2}-\Delta_{3}\right)}\dot{\Omega}_{0}\nonumber \\
 &  & +\frac{d}{dt}\frac{\Delta_{2}E_{3}^{2}-\Delta_{3}E_{2}^{2}}{E_{2}^{2}E_{3}^{2}\left(\Delta_{2}-\Delta_{3}\right)}\dot{\Omega}_{0},\nonumber 
\end{eqnarray}
with $E_{2}^{2}(t)=\Delta_{2}(t)^{2}+4\Omega_{0}(t)^{2}\lambda_{2}^{2}$
and $E_{3}^{2}(t)=\Delta_{3}(t)^{2}+4\Omega_{0}(t)^{2}\lambda_{3}^{2}$,
and where the time-dependence has been left implicit after the first
line. The selection error (Eq. \ref{eq:selectionerror}) for this
pulse is plotted in Fig.~\ref{fig:Selection-error-for} as the solid
red line. For the first and third derivative we obtain

\begin{align}
\Omega(t) & =\Omega_{0}+i\frac{\Delta_{2}E_{3}^{4}-E_{2}^{4}\Delta_{3}}{E_{2}^{2}E_{3}^{2}\left(E_{2}^{2}-E_{3}^{2}\right)}\dot{\Omega}_{0}\label{eq:1stand3rd_3qubits}\\
 & +i\frac{\Delta_{2}E_{3}^{2}-\Delta_{3}E_{2}^{2}}{E_{2}^{2}E_{3}^{2}\left(E_{2}^{2}-E_{3}^{2}\right)}\dddot{\Omega}_{0},\nonumber 
\end{align}
which is plotted in as the dot-dashed orange line. Using the second
and third derivative the solution is

\begin{align}
\Omega(t) & =\Omega_{0}+i\frac{\Delta_{2}\Delta_{3}\left(E_{2}^{2}-E_{3}^{2}\right)}{E_{2}^{2}E_{3}^{4}\Delta_{2}-E_{2}^{4}E_{3}^{2}\Delta_{3}}\dddot{\Omega}_{0}\label{eq:2ndand3rd_3qubits}\\
 & +\frac{d}{dt}\frac{E_{3}^{4}\Delta_{2}-E_{2}^{4}\Delta_{3}}{E_{2}^{2}E_{3}^{4}\Delta_{2}-E_{2}^{4}E_{3}^{2}\Delta_{3}}\dot{\Omega}_{0},\nonumber \\
\nonumber 
\end{align}
which is plotted in dashed. Finally, for the second and fourth derivative
(real) solution we get

\begin{align}
\Omega(t) & =\Omega_{0}+\frac{d}{dt}\left(\frac{1}{E_{2}^{2}}+\frac{1}{E_{3}^{2}}\right)\dot{\Omega}_{0}\label{eq:2ndand4th_3qubits}\\
 & +\frac{d}{dt}\frac{1}{E_{2}^{2}E_{3}^{2}}\dddot{\Omega}_{0},\nonumber 
\end{align}
which does not require compensating the driving qubit (using Eq. \ref{eq:dragdetun-1})
as for the previous pulses. The error for this pulse is plotted in
Fig.~\ref{fig:Selection-error-for} as the dot-dot-dashed black line.
Other such pulses can also be found for other derivative combinations.

$ $

\subsubsection{Quadruplet solutions}

For four qubits ($n=4$), one driven and three affected by crosstalk,
Eqs. \ref{eq:multidrag-1-1} can once be solved using one main control
$\Omega_{0}(t)$ and three auxiliary controls. Here we show the solution
to the equations using the first, second, and third derivatives

\begin{figure}
\centering \includegraphics[width=0.8\columnwidth]{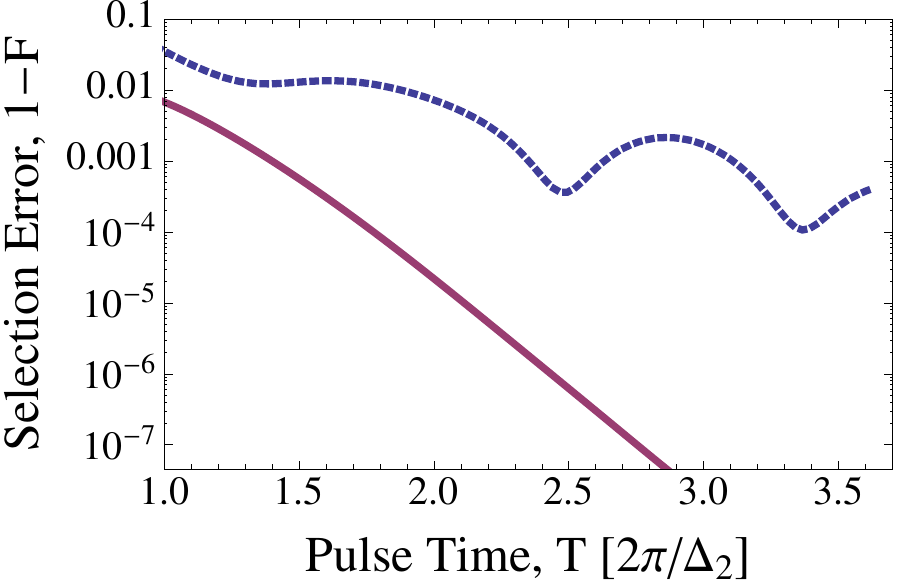} \caption[Selection error for four uncoupled qubits using three derivatives]{Selection error (Eq. \ref{eq:selectionerror}) for four uncoupled
qubits for energy differences $\Delta_{2}$, $\Delta_{3}=1.7\Delta_{2}$,
and $\Delta_{4}=-1.3\Delta_{2}$ from the driven qubit. The dotted
blue line shows the error using standard Gaussian shaping, Eq. \ref{eq:gaussian},
while the solid red line is for the pulse shape given by the first-order
DRAG solution, Eq. \ref{eq:soln_4qubits}. \label{fig:Selection-error-for-1}}
\end{figure}

\begin{widetext}

\begin{eqnarray}
\Omega(t) & = & \Omega_{0}(t)+i\frac{E_{2}^{4}\left(E_{4}^{2}-E_{3}^{2}\right)\Delta_{3}\Delta_{4}+E_{4}^{4}\left(E_{3}^{2}-E_{2}^{2}\right)\Delta_{2}\Delta_{3}+E_{3}^{4}\left(E_{2}^{2}-E_{4}^{2}\right)\Delta_{2}\Delta_{4}}{E_{2}^{2}E_{3}^{2}E_{4}^{2}\left(\left(E_{4}^{2}-E_{3}^{2}\right)\Delta_{2}+\left(E_{2}^{2}-E_{4}^{2}\right)\Delta_{3}+\left(E_{3}^{2}-E_{2}^{2}\right)\Delta_{4}\right)}\dot{\Omega}_{0}(t)\nonumber \\
 &  & +\frac{d}{dt}\frac{\left(E_{3}^{2}E_{4}^{4}-E_{3}^{4}E_{4}^{2}\right)\Delta_{2}+\left(E_{2}^{4}E_{4}^{2}-E_{2}^{2}E_{4}^{4}\right)\Delta_{3}+\left(E_{2}^{2}E_{3}^{4}-E_{3}^{2}E_{2}^{4}\right)\Delta_{4}}{E_{2}^{2}E_{3}^{2}E_{4}^{2}\left(\left(E_{4}^{2}-E_{3}^{2}\right)\Delta_{2}+\left(E_{2}^{2}-E_{4}^{2}\right)\Delta_{3}+\left(E_{3}^{2}-E_{2}^{2}\right)\Delta_{4}\right)}\dot{\Omega}_{0}(t)\label{eq:soln_4qubits}\\
 &  & +i\frac{E_{2}^{2}\left(E_{4}^{2}-E_{3}^{2}\right)\Delta_{3}\Delta_{4}+E_{4}^{2}\left(E_{3}^{2}-E_{2}^{2}\right)\Delta_{2}\Delta_{3}+E_{3}^{2}\left(E_{2}^{2}-E_{4}^{2}\right)\Delta_{2}\Delta_{4}}{E_{2}^{2}E_{3}^{2}E_{4}^{2}\left(\left(E_{4}^{2}-E_{3}^{2}\right)\Delta_{2}+\left(E_{2}^{2}-E_{4}^{2}\right)\Delta_{3}+\left(E_{3}^{2}-E_{2}^{2}\right)\Delta_{4}\right)}\dddot{\Omega}_{0}(t).
\end{eqnarray}
\end{widetext}The selection error vs. gate time for this pulse sequence
(in solid red) and for the Gaussian (in dotted blue) is plotted in
Fig. \ref{fig:Selection-error-for-1}.

\subsection{Discussion}

\begin{figure}
\centering \includegraphics[width=0.8\columnwidth]{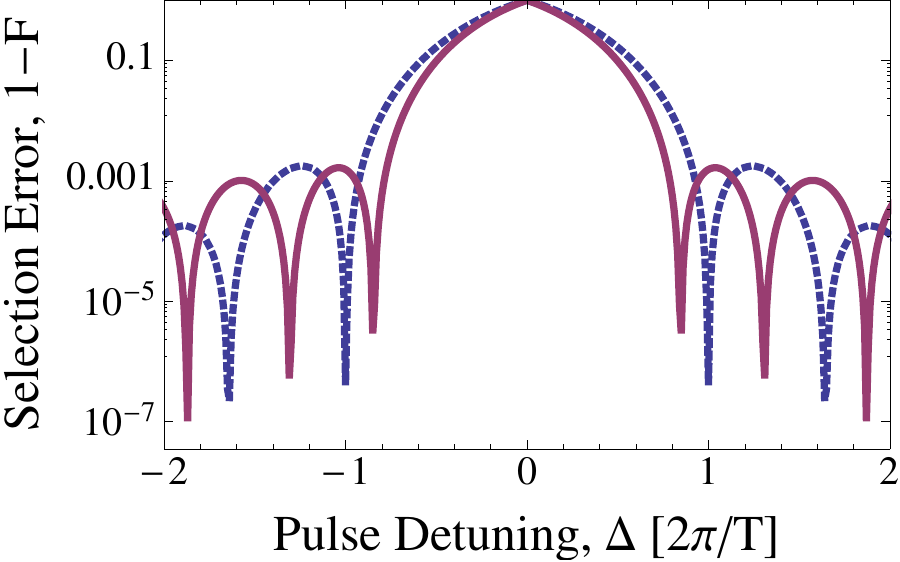}
\caption{Selection error as a function of frequency offset when using a Gaussian
(dotted blue) or Gaussian and its second derivative (solid red). \label{fig:narrowband-1}}
\end{figure}

The advantage of using the expansion which incorporates higher order
derivatives is that each additional order of derivative allows the
removal of (first-order) diabatic error from an unwanted transition, effectively
allowing the adiabatic expansion to be taken into account. Thus, what
we see in Eq. \ref{eq:multidrag-1-1} is that each derivative removes
some portion of the diabatic error so that in aggregate it is fully
removed for each unwanted transition. This is made possible by the
fact that a discrete set of conditions is removed using a discrete
number of control variables. It is also interesting to go back to
look at the case of the continuous excitation spectrum, or one where
the exact position of transitions is not known. As in the semi-classical
case, the effect of adding in derivative terms is to put a hole at
some point in the spectrum, however holes can exist for other reasons,
such as the short time window of the pulse leading to a convolution
of the Gaussian spectrum with a Sinc function. For the second derivative,
the excitation spectrum is shown in Fig. \ref{fig:narrowband-1}.
As before (Fig. \ref{fig:dragspect}B), adding in the second derivative
can have an effect on the bandwidth, here chosen at the cut-off of
0.1\% excitation. Placing the holes appropriately, the DRAG solution
(in red) has a bandwidth \textasciitilde{}25\% narrower than the narrowest
Gaussian pulse (of the same duration, starting at 0) with the same
cut-off. This is consistent with Ref. \cite{Warren84} where second-order
Hermite polynomials are seen to have a similar excitation profile,
though here the location of the holes is engineered\textbf{ }for the
second derivative of the Gaussian with the prefactor $\frac{\ddot{\Omega}}{\Delta^{2}}$.
We see the main reason for the decreased bandwidth is that the area
under the curve is the same as the Gaussian (by conservation of energy),
with weight being moved from the high-excitation region to the tails
in the low-excitation region. Other results for continuous spectra
can be found for other odd or even derivatives, with holes either
one or both sides of the centre frequency, respectively. However,
using higher derivatives comes at a cost, which is that the trial
function (here Eq. \ref{eq:gaussian}) must have its derivatives begin
and end at 0, which effectively increases the bandwidth of the pulse,
and so adding more holes is not necessarily beneficial to an engineered
continuous spectrum.

\section{Connected transitions\label{sec:Connected-transitions}}

Going to higher orders in an expansion of the small parameter may
be necessary for high-precision control as needed, e.g., in implementing
error-correcting protocols. In particular, when transitions are not
disjoint but form a connected graph between energy levels, then the
higher-order effects can be particularly detrimental. The first reason
is that errors on the unwanted transitions will no longer commute
with the working qubit, and errors generated on the working qubit
will typically be more pronounced than were for the other qubits (the
largest energy scale is smaller, $\Omega<\Delta$). The second reason
is that certain resonances may appear between harmonics or sidebands
of the pulse frequency and energy differences between non-nearest-neighbour
energy levels. Finally, as we move to shorter gate times, the errors will become
larger and going to higher orders will be unavoidable.

To demonstrate these detrimental off-resonant effects and their removal
we will consider an anharmonic ladder system. These systems are quite
common and rather ubiquitous in superconducting qubit systems. In
this section we will rederive and then build on the single-transition
removal strategies in \cite{Motzoi09,Gambetta10} which were subsequently
(first) verified in \cite{Chow10,Lucero10}. We show how higher order
effects can be removed in a systematic manner. Note also that in Ref.
\cite{Schutjens13} a combination of disjoint and ladder transitions
was studied with an extension of our approach. 

The general form of the anharmonic ladder in the rotating frame is
the following

\begin{equation}
\begin{split}\hat{H}^{R}(t)= & \sum_{j=1}^{d-1}(j\delta(t)+\Delta_{j})\hat{\Pi}_{j}\\
 & +\sum_{j=1}^{d-1}\lambda_{j-1}\Omega(t)\hat{\sigma}_{j-1,j}^{+}+\mathrm{h.c.},
\end{split}
\label{eq:rotHam-1}
\end{equation}
with $\hat{\sigma}_{j-1,j}^{+}=e^{-i\phi}|j-1\rangle\langle j|$ and
where we assume only one drive at frequency $\omega_{d}\approx\omega_{01}$
(using multiple frequencies allows the system to be solved exactly
\cite{Motzoi2012}). Next we apply the DRAG formalism, Eqs. \ref{eq:dragformalism}.
Going to an interaction frame with respect to the first-order out-of-phase
control (here $\Omega_{1}^{I}(t)=-\frac{\mathrm{Re}\dot{\Omega}(t)}{\Delta}$)
with the transformation

\[
D_{1}=\exp\Bigg(\left(\sum_{j=1}^{d-1}\lambda_{j-1}\int_{0}^{t}\Omega_{1}^{I}(t)dt\right)\hat{\sigma}_{j-1,j}^{+}-\mathrm{h.c.}\Bigg)
\]
gives the interaction Hamiltonian

\begin{equation}
\begin{split}\tilde{H}^{(1)}= & \mathrm{Re}\Omega(t)\left(1+\frac{(4-\lambda^{2})(\mathrm{Re}\Omega(t))^{2}}{2\Delta^{2}}\right)\hat{\sigma}_{0,1}^{+}+\mathrm{h.c.}\\
 & +\left(\delta(t)+\frac{(4-\lambda^{2})(\mathrm{Re}\Omega(t))^{2}}{\Delta}\right)\hat{\Pi}_{1}\\
 & +\left(\Delta+2\delta(t)+\frac{(\lambda^{2}+2)(\mathrm{Re}\Omega(t))^{2}}{\Delta}\right)\hat{\Pi}_{2}\\
 & +\frac{\lambda(\mathrm{Re}\Omega(t))^{2}}{2\Delta}e^{-2i\phi}\hat{\sigma}_{0,2}^{+}+\mathrm{h.c.}+O(\epsilon^{3}).
\end{split}
\end{equation}
where $\hat{\sigma}_{0,2}^{+}=e^{-i2\phi}|0\rangle\langle2|$. In
this frame, it is easy to see there are three errors associated with
the qubit subspace. The selection error is corrected with the off-phase
derivative control $\mathrm{Im}\Omega(t)=-\frac{\dot{\Omega}_{0}(t)}{\Delta}$,
which in this first order is again is exactly the semiclassical result
(Eq. \ref{eq:drag}). The resonance error can be corrected with $\delta(t)=\frac{(\lambda^{2}-4)\Omega_{0}^{2}}{\Delta}$
by either shifting the eigen-energies of the system or by a combination
of phase ramping and frame compensation (see Appendix A). Finally,\textbf{
}this method changes the rotation angle $\theta$ about the rotation
axis ($e^{-i\phi}\sigma_{0,1}^{+}+\mathrm{h.c.}$), which has to be
compensated  by enforcing the area law $\left(\int_{0}^{T}(\Omega_{0}(t)+\frac{(\lambda^{2}-2)(\Omega_{0}(t))^{3}}{\Delta^{2}}\right)dt=\theta$.
To avoid higher order commutator errors, it is even better to satisfy
the condition at all times by renormalizing $\mathrm{Re}\Omega(t)=\Omega_{0}(t)-\frac{(\lambda^{2}-2)(\Omega_{0}(t))^{3}}{\Delta^{2}}$.

\begin{figure}
\centering \includegraphics[width=0.8\columnwidth]{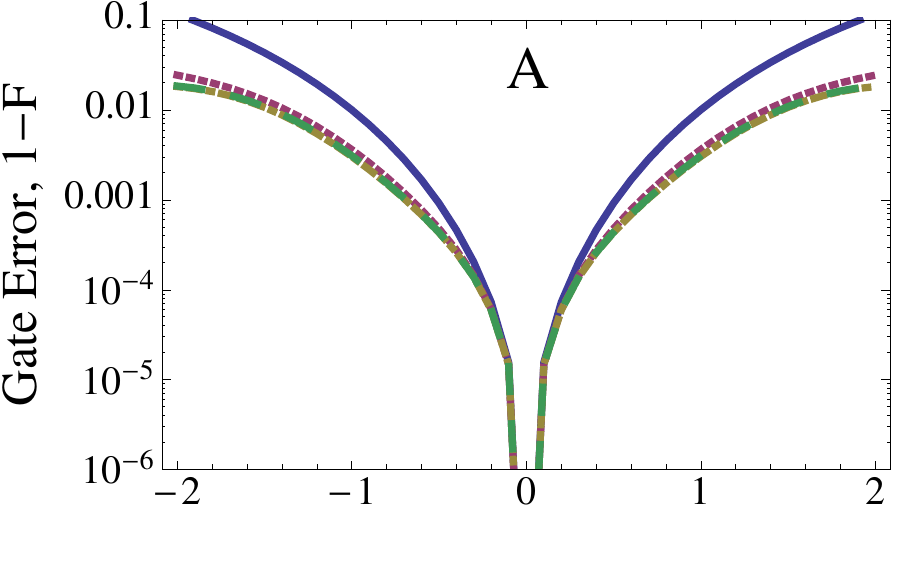}

\includegraphics[width=0.8\columnwidth]{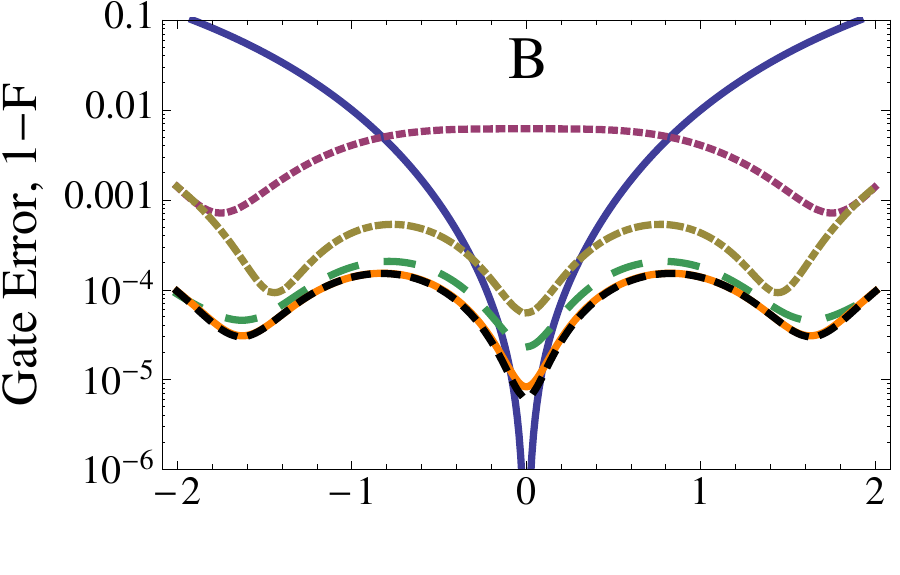}

\includegraphics[width=0.8\columnwidth]{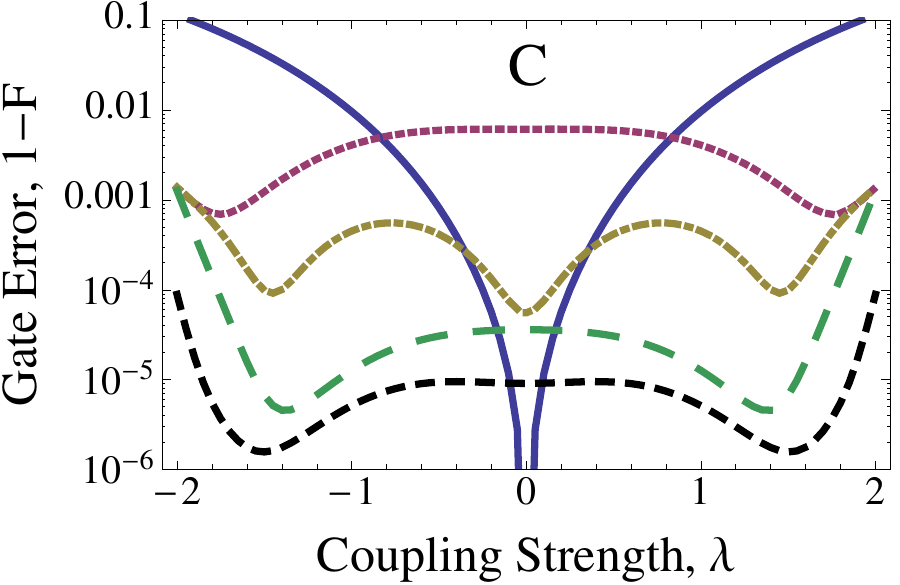}

\caption[Gate error is plotted vs. leakage transition strength $\lambda$ for
solutions to different orders of the adiabatic expansion of a NOT
gate operating on a qutrit]{For a 3-level system driven by a $T=4\pi/\Delta$ with pi-pulse, gate
error is plotted vs. leakage transition strength $\lambda$ for solutions
to the adiabatic expansion to different orders $H^{D(h)}$ correcting
for errors to order $h$. The solid blue line shows the error for
a standard Gaussian, Eq. \ref{eq:gaussian}. Each line under that
gives the error when correcting for the next order in the adiabatic
expansion of the control operators. Frame A gives the solutions to
different orders when no derivative controls are used. Frame B shows
when the $0-1$ transition is removed using a first derivative. Frame
C is the error plotted when an additional perturbative control is
used that includes the second derivative, which enables the removal
of the $0-2$ transition. How the different lines for each order are
calculated is discussed in the text.\label{fig:dragorders}}
\end{figure}

In the next order we see that the error comes from the $\frac{\lambda(\mathrm{Re}\Omega(t))^{2}}{2\Delta}e^{-2i\phi}\hat{\sigma}_{0,2}^{+}$
transition. Now a direct control in the $0-2$ transition is assumed
not present in our system. Therefore we cannot remove this excitation
error with an interaction frame. Instead we must use a composite transformation.
First, we apply the simple next-order diagonalizing transformation
\[
D_{2}=\exp\left(\frac{\lambda\Omega_{0}^{2}(t)}{2\Delta}\hat{\sigma}_{0,2}^{+}/\Delta-\mathrm{h.c.}\right)
\]
with the transformed frame now being

\[
\begin{split}\tilde{H}^{(2)}= & \mathrm{Re}\Omega(t)\left(1+\frac{(4-\lambda^{2})(\mathrm{Re}\Omega(t))^{2}}{2\Delta^{2}}\right)\hat{\sigma}_{0,1}^{+}+\mathrm{h.c.}\\
 & +\left(\frac{\lambda(\mathrm{Re}\Omega(t))^{2}}{2\Delta}-i\frac{\lambda\Omega_{0}\dot{\Omega}_{0}(t)}{\Delta^{2}}-\frac{\lambda\Omega_{0}^{2}(t)}{2\Delta}\right)\hat{\sigma}_{0,2}^{+}+\mathrm{h.c.}\\
 & +\left(\delta(t)+\frac{(4-\lambda^{2})(\mathrm{Re}\Omega(t))^{2}}{\Delta}\right)\hat{\Pi}_{1}\\
 & +\left(\Delta+2\delta(t)+\frac{(\lambda^{2}+2)(\mathrm{Re}\Omega(t))^{2}}{\Delta}\right)\hat{\Pi}_{2}+O(\epsilon^{3}).
\end{split}
\]
but the transformation leaves in the term $+i\dot{\hat{D_{2}}}(t)\hat{D}_{2}^{\dagger}(t)$
term which is of the same order in $\epsilon$ as the term it diagonalizes.
However, in the next order we can obtain a cancelation between the
error terms and the next order auxiliary control. Using the transformation

\[
D_{3}=\exp\left(i\frac{\lambda\Omega_{0}\dot{\Omega}_{0}(t)}{\Delta^{3}}e^{-2i\phi}\hat{\sigma}_{0,2}^{+}/\Delta-\mathrm{h.c.}\right)
\]
we obtain

\begin{equation}
\begin{split}\tilde{H}^{(3)}= & \mathrm{Re}\Omega(t)\left(1+\frac{(4-\lambda^{2})(\mathrm{Re}\Omega(t))^{2}}{2\Delta^{2}}\right)\hat{\sigma}_{0,1}^{+}+\mathrm{h.c.}\\
 & +\left(\frac{\lambda(\mathrm{Re}\Omega(t))^{2}}{2\Delta}-\frac{\lambda(\dot{\Omega}_{0}^{2}+\Omega_{0}\ddot{\Omega}_{0})(t)}{\Delta^{2}}-\frac{\lambda\Omega_{0}^{2}(t)}{2\Delta}\right)\hat{\sigma}_{0,2}^{+}\\
 & +\mathrm{h.c.}+\left(\delta(t)+\frac{(4-\lambda^{2})(\mathrm{Re}\Omega(t))^{2}}{\Delta}\right)\hat{\Pi}_{1}\\
 & +\left(\Delta+2\delta(t)+\frac{(\lambda^{2}+2)(\mathrm{Re}\Omega(t))^{2}}{\Delta}\right)\hat{\Pi}_{2}+O(\epsilon^{3}).
\end{split}
\end{equation}
The most straightforward way to cancel the error is to choose 

\begin{align*}
\mathrm{Re}\Omega(t) & =\Omega_{R}(t)-\frac{(\lambda^{2}-4)(\Omega_{R}(t))^{3}}{2\Delta^{2}}\\
\Omega_{R}(t) & =\sqrt{\frac{\Omega_{0}^{2}(t)}{2}+\frac{(\dot{\Omega}_{0}^{2}+\Omega_{0}\ddot{\Omega}_{0})(t)}{\Delta}}\\
\mathrm{Im}\Omega(t) & =-\frac{\dot{\Omega}_{R}(t)}{\Delta}\\
\delta(t) & =\frac{(\lambda^{2}-4)\Omega_{R}^{2}}{\Delta}
\end{align*}
such that

\begin{equation}
\begin{split}\tilde{H}^{(3)}= & \Omega_{R}(t)\hat{\sigma}_{0,1}^{+}+\mathrm{h.c.}\\
 & +\tilde{\Delta}(t)\hat{\Pi}_{2}+O(\epsilon^{3})
\end{split}
\end{equation}
as required, with $\int_{0}^{R}\Omega_{R}(t)\mathrm{dt}=\theta$ enforcing
the rotation angle. Note that including the second derivative has
allowed us to remove a second undesired transition. The fourth order
can be calculated in a similar way by adding additional perturbations
to the waveform. However, in the fifth order, we will require using
third derivative, which comes from removing the $1-2$ transition
to the next order (which contains a factor in $\ddot{\Omega}$(t)).
Going to higher and higher order, eventually the $0-3$ and $1-3$
transitions will need to be taken into consideration. To demonstrate
the asymptotic bounds derived in Sec. \ref{eq:IBP}, we have plotted
the result of Schrieffer-Wolff diagonalization to multiple orders
when using the standard adiabatic expansion, the interaction frame
for the $1-2$ transition, and the interaction picture for both the
$0-2$ and $1-2$ transitions. Fig. \eqref{fig:dragorders} A, B and
C show these, respectively. The gate fidelity is calculated by

\begin{equation}
F=\frac{1}{2^{n}}|\mathrm{Tr}(U^{\dagger}V)|^{2},\label{eq:gaterror}
\end{equation}
which is the phase-sensitive version of Eq. \ref{eq:selectionerror}.
For each graph, the top blue solid line is the zeroth order and corresponds
to a Gaussian function with area $\pi$. The dotted red line under
it corrects the phase ($\sigma_{z}$) error on the qubit and (for
$B$ and $C$) the selection error via the derivative. The dot-dashed
yellow line under that corrects the second-order rotation angle ($\sigma_{x}$)
error on the qubit. The dashed green line under that corrects error
coming from the $0-2$ transition. The black dashed and solid orange
under those correct the next set of errors. It is clear that using
the {}``interaction frame'' with respect to the auxiliary control(s)
prevents the error from asymptoting as a result of the diabatic error
being undiagonalizable as a direct consequence of the IBP formula,
Eq. \ref{eq:IBP}. In Fig. \eqref{fig:dragorders}C, the proper ordering
of derivatives is used, and we see that indeed each order qualitatively
improves the fidelity compared to the last one.

\section{Selectivity with frequency sidebands\label{sec:rabi}}

As a final example of unwanted off-resonant excitation, we consider
additional frequency components that can be present in a drive and
which we want to suppress. As already mentioned, matrix elements between
non-adjacent levels and simultaneous drives to rotate more than one
transition will have such frequency sidebands. Perhaps the most ubiquitous
occurrence is when applying the rotating-wave approximation to go
from Eq. \ref{eq:lineselec_lab-2} to Eq. \ref{eq:drive_element}
and dropping terms rotating at twice the frequency of the original
drive. That is there are also drive elements of the form 

\begin{eqnarray}
\hat{\Gamma}_{j,k}^{l} & = & \lambda_{j,k}^{l}e^{-i\int_{0}^{t}2\omega_{l}^{d}(t)+\Delta_{jkl}(t)\mathrm{dt}}|j\rangle\langle k|.\label{eq:drive_element-1}
\end{eqnarray}
To characterize the effect of these and similar terms, we move to
the rotating frame, as in Eq. \ref{eq:RotatingFrame}, only rotating
in the opposite direction such that

\begin{align*}
\hat{H}_{R}= & \Omega(t)e^{-i\phi_{l}}\left(e^{+i\int_{0}^{t}2\omega_{d}(t)\mathrm{dt}}+1\right)\hat{\sigma}_{0,1}^{+}+\mathrm{h.c.}\\
 & +2\omega_{d}(t)\ket{1}\bra{1}
\end{align*}
where for clarity we only consider one qubit (on resonance). Note
the fast-oscillating term is actually the term that is on resonance
with the transition and responsible for rotations. The effect of the
off-resonant term can be seen by diagonalizing it with 

\[
D_{1}=\exp\Bigg(\left(\int_{0}^{t}\Omega_{1}^{I}(t)dt\right)\left(1+ce^{+i\int_{0}^{t}2\omega_{d}(t)\mathrm{dt}}\right)\hat{\sigma}_{0,1}^{+}-\mathrm{h.c.}\Bigg)
\]
(with $c$ an arbitrary constant) and choosing $\Omega_{1}^{I}(t)=-c\frac{\dot{\Omega}_{0}}{2\omega_{d}}$
we get

\begin{equation}
\begin{split}\tilde{H}^{(1)}= & \mathrm{Re}\Omega(t)\left(1+e^{+i\int_{0}^{t}2\omega_{d}(t)\mathrm{dt}}\right)\hat{\sigma}_{0,1}^{+}+\\
 & \mathrm{Re}\Omega(t)\left(\frac{1}{2}-c\right)\frac{(\Omega_{0}(t))^{2}}{2\omega_{d}^{2}}\left(1+e^{+i\int_{0}^{t}2\omega_{d}(t)\mathrm{dt}}\right)\hat{\sigma}_{0,1}^{+}\\
 & +i(1-c)\mathrm{Im}\Omega(t)\hat{\sigma}_{0,1}^{+}+\mathrm{h.c.}\\
 & +\left(2\omega_{d}(t)+\left(\frac{1}{2}-c\right)\frac{(\Omega_{0}(t))^{2}}{2\omega_{d}}\right)\ket{1}\bra{1}+O(\epsilon^{3})
\end{split}
\end{equation}
This gives the second order solution (with $c=1$)

\begin{equation}
\begin{split}\Delta_{0,1}(t)=\frac{\Omega_{0}^{2}(t)}{4\omega_{d}}\\
\mathrm{Im}(\Omega(t))=\mathrm{\frac{\dot{\Omega}_{0}(t)}{2\omega_{d}}}\\
\int_{0}^{T}\frac{(\Omega_{0}(t))^{{\rm {eff}}}-(\Omega_{0}(t))^{3}}{4\omega_{d}^{2}}dt & =\theta
\end{split}
\label{eq:rwacorrect}
\end{equation}

\begin{figure}
\centering \includegraphics[width=0.8\columnwidth]{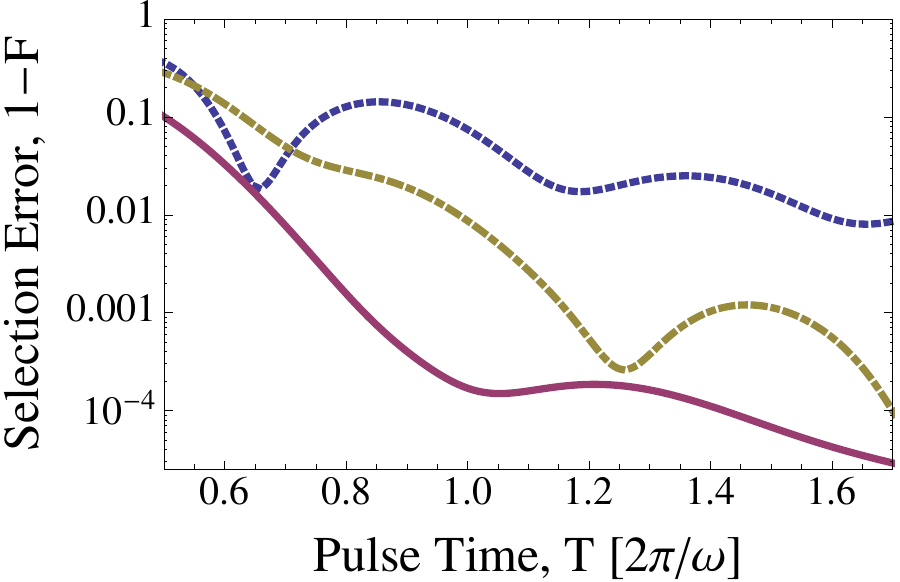} \caption[Error from counter-rotating terms in qubit frame for different analytic
pulses]{Error from counter-rotating terms in qubit frame. Gate error for Gaussian
(dotted blue), Gaussian with derivative (dot-dashed yellow), and Gaussian
with derivative and detuning (solid red) are shown. \label{fig:dragcrt}}
\end{figure}

Fig.~\ref{fig:dragcrt} demonstrates the performance of this strategy.
The dotted blue line shows gate error vs. gate time when using a Gaussian
pulse. The dot-dashed yellow line shows the improvement when optimizing
using only the derivative control and constant drive frequency ($c=\frac{1}{2}$,
see Ref. \cite{Gambetta10}). The solid red line shows the gate error
when both the derivative and detuning are applied, Eq. \eqref{eq:rwacorrect}.
In both cases we see the RWA errors are completely suppressed even
at short times.

\section{Conclusions}

\label{sec:conclusion}

We have shown how derivatives of a pulse shape driving the evolution
of a quantum system can be used to suppress undesired transitions
by introducing auxiliary perturbative controls. In general, multiple
off-resonant transition elements can be removed by using higher derivatives.
The only caveat to these analytic solutions is phase errors need to
be corrected (subsequently) and derivatives need to start and end
at 0 (effectively increasing bandwidth).  To find these solutions to higher orders
a pseudo-adiabatic expansion is performed, in terms of time-instantaneous 
basis functions formed from the derivatives of the trial function. The time-instantaneous
nature of the solutions means they can be expressed easily, even at higher orders, and
that their form does not change with time, only through the time-dependence of the trial
function.  In addition to the computational advantage of being independent of time, the solutions
to higher order are also advantageous in accuracy in comparison to other analytical techniques for which
only low order solutions can be computed, often because the expansions involved are asymptotic.
Here, we have shown that the pseudo-adiabatic expansion does not suffer from asymptotic behaviour
provided higher order derivatives are used to cancel higher-order diabatic errors.

The analytic pulses also motivate numerical and experimental ansatz solutions to other unwanted off-resonant terms
in other (e.g. not fully characterized) physical systems: including higher derivative terms in addition to
amplitude and phase modulation of the shaping function, which ensures
smooth control pulses, may prove to also be a computationally efficient
way to remove undesired terms in a Hamiltonian. 
The derivative-based approach may also
be useful in conjunction with other analytical techniques, such as
dynamical decoupling. In cases where both dynamical decoupling (DD)
  and smooth pulse solutions exist, smooth pulses offer the advantage
  of compatibility with strongly filtered control lines which would
  distort hard DD pulses. Also, the total amount of energy transferred
to the sample is in general lower for smooth than for DD pulses which
is important in cryogenic situations.
 As was pointed out in Ref. \cite{Motzoi09}, the interaction 
picture which is used to motivate DRAG is also used as a first step to the Magnus expansion for DD,
That is, the derivative solutions (Eq.~\eqref{eq:IBP}) are exactly the solution to the decoupled average Hamiltonian.
While we choose to use a small-parameter expansion due to the use of a small perturbative control,
it is possible another expansion such as the Magnus expansion may be more suitable at short times
(where it converges) relative to the inverse of the detuning to the unwanted transition (e.g. for broadband pulses, spin echoes, etc.).

Finally, we have considered three particular classes of physical problems for which derivative removal can
be a successful strategy for combatting off-resonant errors.  These are multiple off-resonant transitions,
multiple higher-order transitions, and sideband transitions.  Worked examples were given to illustrate
the solutions, which were multi-qubit crosstalk, anharmonic ladder transitions, and compensation for
the rotating wave approximation, respectively.  In all three cases, including derivatives of the trial function with
appropriate prefactors was shown to reduce transfer and gate errors by orders of magnitude relative to
conventional pulse shaping techniques.

\begin{acknowledgments}
We thank J. Gambetta, S. Merkel, L. Bishop, and S. Girvin for valuable
discussions. This research was supported by NSERC through the discovery
grants and QuantumWorks. This research was also funded by the Office
of the Director of National Intelligence (ODNI), Intelligence Advanced
Research Projects Activity (IARPA), through the Army Research Office.
 All statements of fact, opinion or conclusions contained herein are
those of the authors and should not be construed as representing the
official views or policies of IARPA, the ODNI, or the U.S. Government. 

\appendix
\end{acknowledgments}

\section{Phase Compensation}

\label{sec:frame}

Using a time-changing drive frequency $\omega_{d}$ such as in Eq.
\eqref{eq:lineselec_lab-2} impacts on the choice of the phase reference
for computational states (or for sequences of non-commuting operations)
. That is, the average frequency $\overline{\omega}_{d}=\int_{0}^{T}\frac{\omega_{d}(t)}{T}$
will differ from the reference given by the nearest qubit, $\omega_{q}.$
This relative phase offset can be compensated by applying a $Z$ operation
that undoes the accumulated phase, or, if such an operator is not
available, by applying a discretely rotating frame with effective
Hamiltonian (after $M$ operations each causing phase offset $ $$U_{\theta}^{Z}$)
given by 
\begin{eqnarray*}
\hat{H}_{\mathrm{{eff}}}(t) & = & (U_{\theta}^{Z})^{M}\hat{H}(U_{-\theta}^{Z})^{M}=U_{M\theta}^{Z}(a(t)\hat{X}+b(t)\hat{Y})U_{-M\theta}^{Z}\\
 & = & (\cos(M\theta)a(t)-\sin(M\theta)b(t))\hat{X}\\
 &  & +(\cos(M\theta)b(t)+\sin(M\theta)a(t))\hat{Y}
\end{eqnarray*}
where $\theta=(\overline{\omega}_{d}-\omega_{q})T.$ If, in addition,
one cannot change the drive frequency time-dependently, one can alternatively
satisfy the requirement with (in the frame rotating at $\overline{\omega}_{d}$) 

\begin{eqnarray*}
\hat{H}_{\mathrm{{eff}}}(t) & = & \exp\left(i\phi_{d}(t)\hat{Z}\right)(a(t)\hat{X}+b(t)\hat{Y})\exp\left(-i\phi_{d}(t)\hat{Z}\right)\\
 & = & \left(\cos\left(\phi_{d}(t)\right)a(t)-\sin\left(\phi_{d}(t)\right)b(t)\right)\hat{X}\\
 &  & +\left(\cos\left(\phi_{d}(t)\right)b(t)+\sin\left(\phi_{d}(t)\right)a(t)\right)\hat{Y}
\end{eqnarray*}
where $ $$\phi_{d}(t)=\int_{0}^{t}(\omega_{d}(t')-\overline{\omega}_{d})dt'$
is the ramped phase. These two phase compensation techniques commute
and can be applied together.

\bibliographystyle{apsrev}

\begin{thebibliography}{45}
\expandafter\ifx\csname natexlab\endcsname\relax\def\natexlab#1{#1}\fi
\expandafter\ifx\csname bibnamefont\endcsname\relax
  \def\bibnamefont#1{#1}\fi
\expandafter\ifx\csname bibfnamefont\endcsname\relax
  \def\bibfnamefont#1{#1}\fi
\expandafter\ifx\csname citenamefont\endcsname\relax
  \def\citenamefont#1{#1}\fi
\expandafter\ifx\csname url\endcsname\relax
  \def\url#1{\texttt{#1}}\fi
\expandafter\ifx\csname urlprefix\endcsname\relax\def\urlprefix{URL }\fi
\providecommand{\bibinfo}[2]{#2}
\providecommand{\eprint}[2][]{\url{#2}}

\bibitem[{\citenamefont{Slichter}(1996)}]{Slichter96}
\bibinfo{author}{\bibfnamefont{C.}~\bibnamefont{Slichter}},
  \emph{\bibinfo{title}{Principles of magnetic resonance}},
  no.~\bibinfo{number}{1} in \bibinfo{series}{Series in Solid State Sciences}
  (\bibinfo{publisher}{Springer}, \bibinfo{address}{Berlin},
  \bibinfo{year}{1996}), \bibinfo{edition}{3rd} ed.

\bibitem[{\citenamefont{Demtr\"oder}(2008)}]{Demtroeder08}
\bibinfo{author}{\bibfnamefont{W.}~\bibnamefont{Demtr\"oder}},
  \emph{\bibinfo{title}{Laser Spectroscopy}}, vol.~\bibinfo{volume}{1}
  (\bibinfo{publisher}{Springer}, \bibinfo{address}{Berlin},
  \bibinfo{year}{2008}).

\bibitem[{\citenamefont{Levitt}(2008)}]{Levitt08}
\bibinfo{author}{\bibfnamefont{M.}~\bibnamefont{Levitt}},
  \emph{\bibinfo{title}{Spin Dynamics. Basics of Nuclear Magnetic Resonance}}
  (\bibinfo{publisher}{Wiley}, \bibinfo{address}{Chichester},
  \bibinfo{year}{2008}), \bibinfo{edition}{2nd} ed.

\bibitem[{\citenamefont{Freeman}(1998)}]{Freeman98}
\bibinfo{author}{\bibfnamefont{R.}~\bibnamefont{Freeman}},
  \emph{\bibinfo{title}{Spin Choreography: Basic Steps in High Resolution NMR}}
  (\bibinfo{publisher}{Oxford University Press}, \bibinfo{address}{New York},
  \bibinfo{year}{1998}).

\bibitem[{\citenamefont{Ernst and and. A.~Wokaun}(1990)}]{Ernst90}
\bibinfo{author}{\bibfnamefont{R.}~\bibnamefont{Ernst}} \bibnamefont{and}
  \bibinfo{author}{\bibfnamefont{G.~B.} \bibnamefont{and. A.~Wokaun}},
  \emph{\bibinfo{title}{Principles of Nuclear Magnetic Resonance in One and Two
  Dimensions}}, International series of monographs on chemistry
  (\bibinfo{publisher}{Oxford University Press}, \bibinfo{address}{Oxford},
  \bibinfo{year}{1990}).

\bibitem[{\citenamefont{Chen et~al.}(2006)\citenamefont{Chen, Church, Englert,
  Henkel, Rohwedder, Scully, and Zubairy}}]{Chen06}
\bibinfo{author}{\bibfnamefont{G.}~\bibnamefont{Chen}},
  \bibinfo{author}{\bibfnamefont{D.}~\bibnamefont{Church}},
  \bibinfo{author}{\bibfnamefont{B.-G.} \bibnamefont{Englert}},
  \bibinfo{author}{\bibfnamefont{C.}~\bibnamefont{Henkel}},
  \bibinfo{author}{\bibfnamefont{B.}~\bibnamefont{Rohwedder}},
  \bibinfo{author}{\bibfnamefont{M.}~\bibnamefont{Scully}}, \bibnamefont{and}
  \bibinfo{author}{\bibfnamefont{M.}~\bibnamefont{Zubairy}},
  \emph{\bibinfo{title}{Quantum Computing Devices: Principles, Designs, and
  Analysis}} (\bibinfo{publisher}{Chapman and Hall/CRC}, \bibinfo{address}{Boca
  Raton}, \bibinfo{year}{2006}).

\bibitem[{\citenamefont{Esteve et~al.}(2004)\citenamefont{Esteve, Raimond, and
  Dalibard}}]{Houches04}
\bibinfo{editor}{\bibfnamefont{D.}~\bibnamefont{Esteve}},
  \bibinfo{editor}{\bibfnamefont{J.-M.} \bibnamefont{Raimond}},
  \bibnamefont{and} \bibinfo{editor}{\bibfnamefont{J.}~\bibnamefont{Dalibard}},
  eds., \emph{\bibinfo{title}{Superconducting qubits and the physics of
  Josephson junctions}}, vol. \bibinfo{volume}{LXXIX} of
  \emph{\bibinfo{series}{Les Houches Session}},
  \bibinfo{organization}{Universite Joseph Fourier}
  (\bibinfo{publisher}{Elsevier}, \bibinfo{address}{Amsterdam},
  \bibinfo{year}{2004}).

\bibitem[{\citenamefont{Nielsen and Chuang}(2000)}]{Nielsen00}
\bibinfo{author}{\bibfnamefont{M.}~\bibnamefont{Nielsen}} \bibnamefont{and}
  \bibinfo{author}{\bibfnamefont{I.}~\bibnamefont{Chuang}},
  \emph{\bibinfo{title}{Quantum Computation and Quantum Information}}
  (\bibinfo{publisher}{Cambridge University Press},
  \bibinfo{address}{Cambridge, UK}, \bibinfo{year}{2000}).

\bibitem[{\citenamefont{Nakahara et~al.}(2006)\citenamefont{Nakahara,
  Kanemitsu, and Salomaa}}]{DiV06}
\bibinfo{editor}{\bibfnamefont{M.}~\bibnamefont{Nakahara}},
  \bibinfo{editor}{\bibfnamefont{S.}~\bibnamefont{Kanemitsu}},
  \bibnamefont{and} \bibinfo{editor}{\bibfnamefont{M.}~\bibnamefont{Salomaa}},
  eds., \emph{\bibinfo{title}{Physical Realizatinons of Quantum Computing: Are
  the DiVincenzo Criteria Fulfilled in 2004?}}
  (\bibinfo{publisher}{WorldScientific}, \bibinfo{address}{Singapore},
  \bibinfo{year}{2006}).

\bibitem[{\citenamefont{Gershenfeld and Chuang}(1997)}]{Gershenfeld97}
\bibinfo{author}{\bibfnamefont{N.}~\bibnamefont{Gershenfeld}} \bibnamefont{and}
  \bibinfo{author}{\bibfnamefont{I.}~\bibnamefont{Chuang}},
  \bibinfo{journal}{Science} \textbf{\bibinfo{volume}{275}},
  \bibinfo{pages}{350} (\bibinfo{year}{1997}).

\bibitem[{\citenamefont{Hughes et~al.}()\citenamefont{Hughes, the ARDA Quantum
  Information~Science, and Panel}}]{qist}
\bibinfo{author}{\bibfnamefont{R.}~\bibnamefont{Hughes}},
  \bibinfo{author}{\bibnamefont{the ARDA Quantum Information~Science}},
  \bibnamefont{and} \bibinfo{author}{\bibfnamefont{T.}~\bibnamefont{Panel}},
  \emph{\bibinfo{title}{A quantum information science and technology roadmap}},
  \bibinfo{howpublished}{available at http://qist.lanl.gov}.

\bibitem[{\citenamefont{Baugh et~al.}(2007)\citenamefont{Baugh, Chamilliard,
  Chandrasekhar, Ditty, Hubbard, Laflamme, Laforest, Moussa, Negrevergne, Silva
  et~al.}}]{Baugh07}
\bibinfo{author}{\bibfnamefont{J.}~\bibnamefont{Baugh}},
  \bibinfo{author}{\bibfnamefont{J.}~\bibnamefont{Chamilliard}},
  \bibinfo{author}{\bibfnamefont{C.}~\bibnamefont{Chandrasekhar}},
  \bibinfo{author}{\bibfnamefont{M.}~\bibnamefont{Ditty}},
  \bibinfo{author}{\bibfnamefont{A.}~\bibnamefont{Hubbard}},
  \bibinfo{author}{\bibfnamefont{R.}~\bibnamefont{Laflamme}},
  \bibinfo{author}{\bibfnamefont{M.}~\bibnamefont{Laforest}},
  \bibinfo{author}{\bibfnamefont{O.}~\bibnamefont{Moussa}},
  \bibinfo{author}{\bibfnamefont{C.}~\bibnamefont{Negrevergne}},
  \bibinfo{author}{\bibfnamefont{M.}~\bibnamefont{Silva}},
  \bibnamefont{et~al.}, \bibinfo{journal}{Phyiscs in Canada}
  \textbf{\bibinfo{volume}{63}}, \bibinfo{pages}{197} (\bibinfo{year}{2007}).

\bibitem[{\citenamefont{Devoret et~al.}()\citenamefont{Devoret, Wallraff, and
  Martinis}}]{Devoret04}
\bibinfo{author}{\bibfnamefont{M.}~\bibnamefont{Devoret}},
  \bibinfo{author}{\bibfnamefont{A.}~\bibnamefont{Wallraff}}, \bibnamefont{and}
  \bibinfo{author}{\bibfnamefont{J.}~\bibnamefont{Martinis}},
  \bibinfo{note}{cond-mat/0411174}.

\bibitem[{\citenamefont{Clarke and Wilhelm}(2008)}]{Insight}
\bibinfo{author}{\bibfnamefont{J.}~\bibnamefont{Clarke}} \bibnamefont{and}
  \bibinfo{author}{\bibfnamefont{F.}~\bibnamefont{Wilhelm}},
  \bibinfo{journal}{Nature} \textbf{\bibinfo{volume}{453}},
  \bibinfo{pages}{1031} (\bibinfo{year}{2008}).

\bibitem[{\citenamefont{Makhlin et~al.}(2001)\citenamefont{Makhlin, Sch\"on,
  and Shnirman}}]{Makhlin01}
\bibinfo{author}{\bibfnamefont{Y.}~\bibnamefont{Makhlin}},
  \bibinfo{author}{\bibfnamefont{G.}~\bibnamefont{Sch\"on}}, \bibnamefont{and}
  \bibinfo{author}{\bibfnamefont{A.}~\bibnamefont{Shnirman}},
  \bibinfo{journal}{Rev. Mod. Phys.} \textbf{\bibinfo{volume}{73}},
  \bibinfo{pages}{357} (\bibinfo{year}{2001}).

\bibitem[{\citenamefont{Schoelkopf and Girvin}(2008)}]{Schoelkopf08}
\bibinfo{author}{\bibfnamefont{R.}~\bibnamefont{Schoelkopf}} \bibnamefont{and}
  \bibinfo{author}{\bibfnamefont{S.}~\bibnamefont{Girvin}},
  \bibinfo{journal}{Nature} \textbf{\bibinfo{volume}{451}},
  \bibinfo{pages}{664} (\bibinfo{year}{2008}).

\bibitem[{\citenamefont{You and Nori}(2005)}]{You05b}
\bibinfo{author}{\bibfnamefont{J.}~\bibnamefont{You}} \bibnamefont{and}
  \bibinfo{author}{\bibfnamefont{F.}~\bibnamefont{Nori}},
  \bibinfo{journal}{Phys. Today} \textbf{\bibinfo{volume}{58}},
  \bibinfo{pages}{42} (\bibinfo{year}{2005}).

\bibitem[{\citenamefont{Blais et~al.}(2007)\citenamefont{Blais, Gambetta,
  Wallraff, Schuster, Girvin, Devoret, and Schoelkopf}}]{Blais07}
\bibinfo{author}{\bibfnamefont{A.}~\bibnamefont{Blais}},
  \bibinfo{author}{\bibfnamefont{J.}~\bibnamefont{Gambetta}},
  \bibinfo{author}{\bibfnamefont{A.}~\bibnamefont{Wallraff}},
  \bibinfo{author}{\bibfnamefont{D.}~\bibnamefont{Schuster}},
  \bibinfo{author}{\bibfnamefont{S.}~\bibnamefont{Girvin}},
  \bibinfo{author}{\bibfnamefont{M.}~\bibnamefont{Devoret}}, \bibnamefont{and}
  \bibinfo{author}{\bibfnamefont{R.}~\bibnamefont{Schoelkopf}},
  \bibinfo{journal}{Phys. Rev. A} \textbf{\bibinfo{volume}{75}},
  \bibinfo{pages}{032329} (\bibinfo{year}{2007}).

\bibitem[{\citenamefont{Hofheinz et~al.}(2009)\citenamefont{Hofheinz, Wang,
  Ansmann, Bialczak, Lucero, Neeley, O'Connell, Sank, Wenner, Martinis
  et~al.}}]{Hofheinz09}
\bibinfo{author}{\bibfnamefont{M.}~\bibnamefont{Hofheinz}},
  \bibinfo{author}{\bibfnamefont{H.}~\bibnamefont{Wang}},
  \bibinfo{author}{\bibfnamefont{M.}~\bibnamefont{Ansmann}},
  \bibinfo{author}{\bibfnamefont{R.}~\bibnamefont{Bialczak}},
  \bibinfo{author}{\bibfnamefont{E.}~\bibnamefont{Lucero}},
  \bibinfo{author}{\bibfnamefont{M.}~\bibnamefont{Neeley}},
  \bibinfo{author}{\bibfnamefont{A.}~\bibnamefont{O'Connell}},
  \bibinfo{author}{\bibfnamefont{D.}~\bibnamefont{Sank}},
  \bibinfo{author}{\bibfnamefont{J.}~\bibnamefont{Wenner}},
  \bibinfo{author}{\bibfnamefont{J.}~\bibnamefont{Martinis}},
  \bibnamefont{et~al.}, \bibinfo{journal}{Nature}
  \textbf{\bibinfo{volume}{459}}, \bibinfo{pages}{546} (\bibinfo{year}{2009}).

\bibitem[{\citenamefont{Majer et~al.}(2007)\citenamefont{Majer, Chow, Gambetta,
  Koch, Johnson, Schreier, Frunzio, Schuster, Houck, Wallraff
  et~al.}}]{Majer07}
\bibinfo{author}{\bibfnamefont{J.}~\bibnamefont{Majer}},
  \bibinfo{author}{\bibfnamefont{J.}~\bibnamefont{Chow}},
  \bibinfo{author}{\bibfnamefont{J.}~\bibnamefont{Gambetta}},
  \bibinfo{author}{\bibfnamefont{J.}~\bibnamefont{Koch}},
  \bibinfo{author}{\bibfnamefont{B.}~\bibnamefont{Johnson}},
  \bibinfo{author}{\bibfnamefont{J.}~\bibnamefont{Schreier}},
  \bibinfo{author}{\bibfnamefont{L.}~\bibnamefont{Frunzio}},
  \bibinfo{author}{\bibfnamefont{D.}~\bibnamefont{Schuster}},
  \bibinfo{author}{\bibfnamefont{A.}~\bibnamefont{Houck}},
  \bibinfo{author}{\bibfnamefont{A.}~\bibnamefont{Wallraff}},
  \bibnamefont{et~al.}, \bibinfo{journal}{Nature}
  \textbf{\bibinfo{volume}{449}}, \bibinfo{pages}{443} (\bibinfo{year}{2007}).

\bibitem[{\citenamefont{Plantenberg et~al.}(2007)\citenamefont{Plantenberg,
  de~Groot, Harmans, and Mooij}}]{Plantenberg07}
\bibinfo{author}{\bibfnamefont{J.~H.} \bibnamefont{Plantenberg}},
  \bibinfo{author}{\bibfnamefont{P.~C.} \bibnamefont{de~Groot}},
  \bibinfo{author}{\bibfnamefont{C.~J. P.~M.} \bibnamefont{Harmans}},
  \bibnamefont{and} \bibinfo{author}{\bibfnamefont{J.~E.} \bibnamefont{Mooij}},
  \bibinfo{journal}{Nature} \textbf{\bibinfo{volume}{447}},
  \bibinfo{pages}{836} (\bibinfo{year}{2007}).

\bibitem[{\citenamefont{Sillanp\"a\"a et~al.}(2007)\citenamefont{Sillanp\"a\"a,
  Park, and Simmonds}}]{Sillanpaa07}
\bibinfo{author}{\bibfnamefont{M.}~\bibnamefont{Sillanp\"a\"a}},
  \bibinfo{author}{\bibfnamefont{J.}~\bibnamefont{Park}}, \bibnamefont{and}
  \bibinfo{author}{\bibfnamefont{R.}~\bibnamefont{Simmonds}},
  \bibinfo{journal}{Nature} \textbf{\bibinfo{volume}{449}},
  \bibinfo{pages}{438} (\bibinfo{year}{2007}).

\bibitem[{\citenamefont{Galiautdinov et~al.}()\citenamefont{Galiautdinov,
  Korotkov, and Martinis}}]{Galiautdinov11}
\bibinfo{author}{\bibfnamefont{A.}~\bibnamefont{Galiautdinov}},
  \bibinfo{author}{\bibfnamefont{A.}~\bibnamefont{Korotkov}}, \bibnamefont{and}
  \bibinfo{author}{\bibfnamefont{J.}~\bibnamefont{Martinis}},
  \bibinfo{note}{arXiv:1105.3997}.

\bibitem[{\citenamefont{Loy}(1974)}]{Loy74}
\bibinfo{author}{\bibfnamefont{M.}~\bibnamefont{Loy}}, \bibinfo{journal}{Phys.
  Rev. Lett.} \textbf{\bibinfo{volume}{32}}, \bibinfo{pages}{814}
  (\bibinfo{year}{1974}).

\bibitem[{\citenamefont{Unanyan et~al.}(1997)\citenamefont{Unanyan, Yatsenko,
  Bergmann, and Shore}}]{Unanyanc1997}
\bibinfo{author}{\bibfnamefont{R.~G.} \bibnamefont{Unanyan}},
  \bibinfo{author}{\bibfnamefont{L.~P.} \bibnamefont{Yatsenko}},
  \bibinfo{author}{\bibfnamefont{K.}~\bibnamefont{Bergmann}}, \bibnamefont{and}
  \bibinfo{author}{\bibfnamefont{B.~W.} \bibnamefont{Shore}},
  \bibinfo{journal}{Opt. Commun.} \textbf{\bibinfo{volume}{139}},
  \bibinfo{pages}{48} (\bibinfo{year}{1997}).

\bibitem[{\citenamefont{Demirplak and Rice}(2003)}]{Demirplak2003}
\bibinfo{author}{\bibfnamefont{M.}~\bibnamefont{Demirplak}} \bibnamefont{and}
  \bibinfo{author}{\bibfnamefont{S.~A.} \bibnamefont{Rice}},
  \bibinfo{journal}{J. Phys. Chem. A} \textbf{\bibinfo{volume}{107}},
  \bibinfo{pages}{9937} (\bibinfo{year}{2003}).

\bibitem[{\citenamefont{Motzoi et~al.}(2009)\citenamefont{Motzoi, Gambetta,
  Rebentrost, and Wilhelm}}]{Motzoi09}
\bibinfo{author}{\bibfnamefont{F.}~\bibnamefont{Motzoi}},
  \bibinfo{author}{\bibfnamefont{J.}~\bibnamefont{Gambetta}},
  \bibinfo{author}{\bibfnamefont{P.}~\bibnamefont{Rebentrost}},
  \bibnamefont{and} \bibinfo{author}{\bibfnamefont{F.}~\bibnamefont{Wilhelm}},
  \bibinfo{journal}{Phys. Rev. Lett} \textbf{\bibinfo{volume}{103}},
  \bibinfo{pages}{110501} (\bibinfo{year}{2009}).

\bibitem[{\citenamefont{Berry}(2009)}]{Berry09}
\bibinfo{author}{\bibfnamefont{M.}~\bibnamefont{Berry}}, \bibinfo{journal}{J.
  Phys. A: Math. Theor.} \textbf{\bibinfo{volume}{42}}, \bibinfo{pages}{365303}
  (\bibinfo{year}{2009}).

\bibitem[{\citenamefont{Gambetta et~al.}(2011)\citenamefont{Gambetta, Motzoi,
  Merkel, and Wilhelm}}]{Gambetta10}
\bibinfo{author}{\bibfnamefont{J.}~\bibnamefont{Gambetta}},
  \bibinfo{author}{\bibfnamefont{F.}~\bibnamefont{Motzoi}},
  \bibinfo{author}{\bibfnamefont{S.}~\bibnamefont{Merkel}}, \bibnamefont{and}
  \bibinfo{author}{\bibfnamefont{F.}~\bibnamefont{Wilhelm}},
  \bibinfo{journal}{Phys. Rev. A} \textbf{\bibinfo{volume}{83}},
  \bibinfo{pages}{012308} (\bibinfo{year}{2011}).


\bibitem[{\citenamefont{Motzoi}(2012)}]{Motzoi2012}
\bibinfo{author}{\bibfnamefont{F.}~\bibnamefont{Motzoi}}, Ph.D. thesis,
  \bibinfo{school}{University of Waterloo} (\bibinfo{year}{2012}).

\bibitem[{\citenamefont{Guerin et~al.}(2011)\citenamefont{Guerin, Hakobyan, and
  Jauslin}}]{Guerin11}
\bibinfo{author}{\bibfnamefont{S.}~\bibnamefont{Guerin}},
  \bibinfo{author}{\bibfnamefont{V.}~\bibnamefont{Hakobyan}}, \bibnamefont{and}
  \bibinfo{author}{\bibfnamefont{H.}~\bibnamefont{Jauslin}},
  \bibinfo{journal}{Phys. Rev. A} \textbf{\bibinfo{volume}{84}},
  \bibinfo{pages}{013423} (\bibinfo{year}{2011}).

\bibitem[{\citenamefont{Torrontegui et~al.}()\citenamefont{Torrontegui, Ibanez,
  Martinez-Garaot, Modugno, del Campo, Guery-Odelin, Ruschhaupt, Chen, and
  Muga}}]{Torrontegui12}
\bibinfo{author}{\bibfnamefont{E.}~\bibnamefont{Torrontegui}},
  \bibinfo{author}{\bibfnamefont{S.}~\bibnamefont{Ibanez}},
  \bibinfo{author}{\bibfnamefont{S.}~\bibnamefont{Martinez-Garaot}},
  \bibinfo{author}{\bibfnamefont{M.}~\bibnamefont{Modugno}},
  \bibinfo{author}{\bibfnamefont{A.}~\bibnamefont{del Campo}},
  \bibinfo{author}{\bibfnamefont{D.}~\bibnamefont{Guery-Odelin}},
  \bibinfo{author}{\bibfnamefont{A.}~\bibnamefont{Ruschhaupt}},
  \bibinfo{author}{\bibfnamefont{X.}~\bibnamefont{Chen}}, \bibnamefont{and}
  \bibinfo{author}{\bibfnamefont{J.~G.} \bibnamefont{Muga}},
  \bibinfo{note}{arXiv:1212.6343}.

\bibitem[{\citenamefont{del Campo et~al.}(2012)\citenamefont{del Campo, Rams,
  and M.}}]{delCampo2012}
\bibinfo{author}{\bibfnamefont{A.}~\bibnamefont{del Campo}},
  \bibinfo{author}{\bibfnamefont{M.}~\bibnamefont{Rams}}, \bibnamefont{and}
  \bibinfo{author}{\bibfnamefont{W.~H.} \bibnamefont{M.},
  \bibfnamefont{Zurek}}, \bibinfo{journal}{Phys. Rev. Lett.}
  \textbf{\bibinfo{volume}{109}}, \bibinfo{pages}{115703}
  (\bibinfo{year}{2012}).

\bibitem[{\citenamefont{Chow et~al.}(2010)\citenamefont{Chow, DiCarlo,
  Gambetta, Motzoi, Frunzio, Girvin, and Schoelkopf}}]{Chow10}
\bibinfo{author}{\bibfnamefont{J.}~\bibnamefont{Chow}},
  \bibinfo{author}{\bibfnamefont{L.}~\bibnamefont{DiCarlo}},
  \bibinfo{author}{\bibfnamefont{J.}~\bibnamefont{Gambetta}},
  \bibinfo{author}{\bibfnamefont{F.}~\bibnamefont{Motzoi}},
  \bibinfo{author}{\bibfnamefont{L.}~\bibnamefont{Frunzio}},
  \bibinfo{author}{\bibfnamefont{S.}~\bibnamefont{Girvin}}, \bibnamefont{and}
  \bibinfo{author}{\bibfnamefont{R.}~\bibnamefont{Schoelkopf}},
  \bibinfo{journal}{Phys. Rev. A} \textbf{\bibinfo{volume}{82}},
  \bibinfo{pages}{040305(R)} (\bibinfo{year}{2010}).

\bibitem[{\citenamefont{Lucero et~al.}(2010)\citenamefont{Lucero, Kelly,
  Bialczak, Lenander, Mariantoni, Neeley, O'Connell, Sank, Wang, Weides
  et~al.}}]{Lucero10}
\bibinfo{author}{\bibfnamefont{E.}~\bibnamefont{Lucero}},
  \bibinfo{author}{\bibfnamefont{J.}~\bibnamefont{Kelly}},
  \bibinfo{author}{\bibfnamefont{R.}~\bibnamefont{Bialczak}},
  \bibinfo{author}{\bibfnamefont{M.}~\bibnamefont{Lenander}},
  \bibinfo{author}{\bibfnamefont{M.}~\bibnamefont{Mariantoni}},
  \bibinfo{author}{\bibfnamefont{M.}~\bibnamefont{Neeley}},
  \bibinfo{author}{\bibfnamefont{A.}~\bibnamefont{O'Connell}},
  \bibinfo{author}{\bibfnamefont{D.}~\bibnamefont{Sank}},
  \bibinfo{author}{\bibfnamefont{H.}~\bibnamefont{Wang}},
  \bibinfo{author}{\bibfnamefont{M.}~\bibnamefont{Weides}},
  \bibnamefont{et~al.}, \bibinfo{journal}{Phys. Rev. A}
  \textbf{\bibinfo{volume}{82}}, \bibinfo{pages}{042339}
  (\bibinfo{year}{2010}).

\bibitem[{\citenamefont{Hoult}(1979)}]{Hoult79}
\bibinfo{author}{\bibfnamefont{D.}~\bibnamefont{Hoult}}, \bibinfo{journal}{J.
  Magn. Res.} \textbf{\bibinfo{volume}{35}}, \bibinfo{pages}{69}
  (\bibinfo{year}{1979}).

\bibitem[{\citenamefont{Vold et~al.}(1968)\citenamefont{Vold, Waugh, Klein, and
  Phelps}}]{Vold68}
\bibinfo{author}{\bibfnamefont{R.}~\bibnamefont{Vold}},
  \bibinfo{author}{\bibfnamefont{J.}~\bibnamefont{Waugh}},
  \bibinfo{author}{\bibfnamefont{M.}~\bibnamefont{Klein}}, \bibnamefont{and}
  \bibinfo{author}{\bibfnamefont{D.}~\bibnamefont{Phelps}},
  \bibinfo{journal}{J. Chem. Phys.} \textbf{\bibinfo{volume}{43}},
  \bibinfo{pages}{3831} (\bibinfo{year}{1968}).

\bibitem[{\citenamefont{Warren}(1984)}]{Warren84}
\bibinfo{author}{\bibfnamefont{W.}~\bibnamefont{Warren}}, \bibinfo{journal}{J.
  Chem. Phys.} \textbf{\bibinfo{volume}{81}}, \bibinfo{pages}{5437}
  (\bibinfo{year}{1984}).

\bibitem[{\citenamefont{Haeberlen and Waugh}(1968)}]{Haeberlen68}
\bibinfo{author}{\bibfnamefont{U.}~\bibnamefont{Haeberlen}} \bibnamefont{and}
  \bibinfo{author}{\bibfnamefont{J.~S.} \bibnamefont{Waugh}},
  \bibinfo{journal}{Phys. Rev.} \textbf{\bibinfo{volume}{175}},
  \bibinfo{pages}{453} (\bibinfo{year}{1968}).

\bibitem[{\citenamefont{Bauer et~al.}(1984)\citenamefont{Bauer, Freeman,
  Frenkiel, Keeler, and Shaka}}]{Bauer1984}
\bibinfo{author}{\bibfnamefont{C.}~\bibnamefont{Bauer}},
  \bibinfo{author}{\bibfnamefont{R.}~\bibnamefont{Freeman}},
  \bibinfo{author}{\bibfnamefont{T.}~\bibnamefont{Frenkiel}},
  \bibinfo{author}{\bibfnamefont{J.}~\bibnamefont{Keeler}}, \bibnamefont{and}
  \bibinfo{author}{\bibfnamefont{A.~J.} \bibnamefont{Shaka}},
  \bibinfo{journal}{J. Magn. Reson.} \textbf{\bibinfo{volume}{58}},
  \bibinfo{pages}{442} (\bibinfo{year}{1984}).

\bibitem[{\citenamefont{Steffen et~al.}(2003)\citenamefont{Steffen, Martinis,
  and Chuang}}]{Steffen2003}
\bibinfo{author}{\bibfnamefont{M.}~\bibnamefont{Steffen}},
  \bibinfo{author}{\bibfnamefont{J.~M.} \bibnamefont{Martinis}},
  \bibnamefont{and} \bibinfo{author}{\bibfnamefont{I.~L.}
  \bibnamefont{Chuang}}, \bibinfo{journal}{Phys. Rev. B}
  \textbf{\bibinfo{volume}{68}}, \bibinfo{pages}{224518}
  (\bibinfo{year}{2003}).

\bibitem[{\citenamefont{Lim and Berry}(1991)}]{Lim1991}
\bibinfo{author}{\bibfnamefont{R.}~\bibnamefont{Lim}} \bibnamefont{and}
  \bibinfo{author}{\bibfnamefont{M.~V.} \bibnamefont{Berry}},
  \bibinfo{journal}{J. Phys. A: Math} \textbf{\bibinfo{volume}{24}},
  \bibinfo{pages}{3255} (\bibinfo{year}{1991}).

\bibitem[{\citenamefont{Deschamps et~al.}(2008)\citenamefont{Deschamps,
  Kervern, Massiot, Pintacuda, Emsley, and Grandinetti}}]{Deschamps2008}
\bibinfo{author}{\bibfnamefont{M.}~\bibnamefont{Deschamps}},
  \bibinfo{author}{\bibfnamefont{G.}~\bibnamefont{Kervern}},
  \bibinfo{author}{\bibfnamefont{D.}~\bibnamefont{Massiot}},
  \bibinfo{author}{\bibfnamefont{G.}~\bibnamefont{Pintacuda}},
  \bibinfo{author}{\bibfnamefont{L.}~\bibnamefont{Emsley}}, \bibnamefont{and}
  \bibinfo{author}{\bibfnamefont{P.~J.} \bibnamefont{Grandinetti}},
  \bibinfo{journal}{J. Chem. Phys.} \textbf{\bibinfo{volume}{129}},
  \bibinfo{pages}{204110} (\bibinfo{year}{2008}).

\bibitem[{\citenamefont{R. and A.}(1966)}]{Schrieffer1966}
\bibinfo{author}{\bibfnamefont{S.~J.} \bibnamefont{R.}} \bibnamefont{and}
  \bibinfo{author}{\bibfnamefont{W.~P.} \bibnamefont{A.}},
  \bibinfo{journal}{Phys. Rev.} \textbf{\bibinfo{volume}{2}},
  \bibinfo{pages}{491} (\bibinfo{year}{1966}).

\bibitem[{\citenamefont{Schutjens et~al.}()\citenamefont{Schutjens, Dagga,
  Egger, and Wilhelm}}]{Schutjens13}
\bibinfo{author}{\bibfnamefont{R.}~\bibnamefont{Schutjens}},
  \bibinfo{author}{\bibfnamefont{F.~A.} \bibnamefont{Dagga}},
  \bibinfo{author}{\bibfnamefont{D.}~\bibnamefont{Egger}}, \bibnamefont{and}
  \bibinfo{author}{\bibfnamefont{F.}~\bibnamefont{Wilhelm}},
  \bibinfo{note}{arXiv:1306.2279}.


\end{thebibliography}

\end{document}